\pdfoutput=1
\documentclass[aip,jcp, reprint,floatfix,a4paper]{revtex4-1}  
\usepackage{cmap}
\usepackage[T1]{fontenc}
\usepackage[Euler]{upgreek}
\usepackage{times}
\usepackage{helvet}
\usepackage{fixmath}
\DeclareSymbolFont{UPM}{U}{eur}{m}{n}
\DeclareMathSymbol{\partial}{0}{UPM}{"40}
\usepackage{graphicx}
\usepackage{amsmath}
\usepackage{geometry}
\usepackage{comment}
\usepackage[stretch=15,shrink=15,step=1]{microtype}
\SetProtrusion{encoding={*},family={*},series={*},size={6,7}}
              {1={ ,550},2={ ,300},3={ ,300},4={ ,300},5={ ,300},
               6={ ,300},7={ ,400},8={ ,300},9={ ,300},0={ ,300}}
\usepackage[nodayofweek]{datetime}
\newdateformat{myDate}{\THEDAY\ \monthname[\THEMONTH] \THEYEAR}
\DeclareMathOperator{\erfc}{erfc}

\DeclareMathOperator{\sgn}{sgn}
\DeclareMathOperator{\pbc}{pbc}
\DeclareMathOperator{\round}{round}
\usepackage{bm}
\usepackage{placeins}
\usepackage{csquotes}
\usepackage{mathtools}
\usepackage{bbm}

\usepackage{hyperref}
\hypersetup{
        colorlinks=true,
        citecolor=black,
        filecolor=black,
        linkcolor=black,
        urlcolor=black
}

\begin{document}

\preprint{}

\title[]{Non-equilibrium simulations of thermally
induced electric fields in water.}

\author{P. Wirnsberger}
\affiliation{
{\footnotesize{Department of Chemistry, University of Cambridge, Cambridge CB2 1EW,
United Kingdom.}}%
}%

\author{D. Fijan}
\affiliation{
{\footnotesize{Department of Chemistry, University of Oxford, Oxford OX1 3QZ,
United Kingdom.}} \\
}%

\author{A. \v{S}ari\'c}%
\affiliation{
{\footnotesize{Department of Chemistry, University of Cambridge, Cambridge CB2
1EW, United Kingdom.}}%
}
\affiliation{
{\footnotesize{Department of Physics and Astronomy, Institute for
the Physics of Living Systems, University College London, WC1E 6BT, United
Kingdom.}}%
}%

\author{M. Neumann}
\affiliation{%
{\footnotesize{Faculty of Physics, University of Vienna, 1090 Vienna, Austria.}}
\\
}%

\author{C. Dellago}
\affiliation{%
{\footnotesize{Faculty of Physics, University of Vienna, 1090 Vienna, Austria.}}
\\
}%

\author{D. Frenkel}%
\affiliation{
{\footnotesize{Department of Chemistry, University of Cambridge, Cambridge CB2
1EW, United Kingdom.}}%
}%

\date{\today}
\begin{abstract}
Using non-equilibrium molecular dynamics simulations, it has been
recently demonstrated that water molecules align in response to an imposed temperature gradient, 
resulting in an effective electric field. Here, we investigate how thermally induced fields 
depend on the underlying treatment of long-ranged interactions. For the
short-ranged Wolf method and Ewald summation, we find the peak strength of the
field to range between $2 \times 10^7$ and $5 \times 10^7~\text{V/m}$ for a temperature gradient of $5.2~\text{K/\AA}$.
Our value for the Wolf method is therefore an order of magnitude lower than the
literature value [J. Chem. Phys. \textbf{139}, 014504 (2013) and \textbf{143},
036101 (2015)]. We show that this discrepancy can be traced back to the use of
an incorrect kernel in the calculation of the electrostatic field. More
seriously, we find that the Wolf method fails to predict correct molecular
orientations, resulting in dipole densities with opposite sign to those computed using Ewald summation. 
By considering two different multipole expansions, we show that, for inhomogeneous polarisations, 
the quadrupole contribution can be significant and even outweigh the dipole contribution to the field. 
Finally, we propose a more accurate way of calculating the electrostatic potential and the field. 
In particular, we show that averaging the microscopic field analytically to obtain the macroscopic 
Maxwell field reduces the error bars by up to an order of magnitude. As a consequence, the simulation 
times required to reach a given statistical accuracy decrease by up to two orders of magnitude.
\end{abstract}


\maketitle

\section{\label{sec:intro}Introduction}
A wide range of phenomena in physics, biology, chemistry and materials science
are caused by strong spatial variations in thermodynamic quantities, such as
pressure or temperature, on a microscopic scale.
Some of these effects are related to temperature gradients which may, for
instance, be induced by ultrasonic insonation~\cite{Doktycz1990} or heated nanoparticles~\cite{Govorov2007}. 
The Peltier effect as well as the Soret
effect both fall in this category~\cite{Bresme2008}.
Another effect, which has received considerable attention recently, is the
thermo-polarisation effect~\cite{Bresme2008,Muscatello2011a,Romer2012,Armstrong2013, Armstrong2013a}.
Using non-equilibrium molecular dynamics (NEMD) simulations, Bresme and
co-workers demonstrated that water molecules align in response to an imposed
temperature gradient, leading to electrostatic fields as high as
$10^8$~V/m for gradients of 5~K/\AA~\cite{Armstrong2013, Armstrong2013a}.
In addition, they were able to confirm that the electric field scales linearly
with the temperature
gradient~\cite{Muscatello2011a,Armstrong2013,Armstrong2013a} in 
accordance with the theoretical predictions of non-equilibrium thermodynamics
(NET)~\cite{Groot2013}.

In molecular simulations, Coulomb
interactions are regularly treated via Ewald
summation~\cite{Ewald1921} (including approximations to it) or a
form of truncated interactions~\cite{Fennell2006}. In most studies on the thermo-polarisation 
effect~\cite{Bresme2008,Muscatello2011a, Armstrong2013, Armstrong2013a}, 
electrostatic interactions were handled with the
truncated, short-ranged Wolf method~\cite{Wolf1999}.
It was argued that Ewald summation can introduce
artifacts, which can be avoided by using the short-ranged
method~\cite{Bresme2008}.
Very recently, however, Bresme and co-workers found that the
Wolf method overestimates the induced electric field in a spherical droplet of
water by an order of magnitude as compared to Ewald
summation~\cite{Armstrong2015}.

The Wolf method and other short-ranged
methods~\cite{Steinbach1994, Zahn2002, Wu2005, Fennell2006,
Elvira2014a, Fukuda2011, Fukuda2013, Lamichhane2014a,
Chen2006a,Fanourgakis2015} are attractive because they achieve linear scaling with the number 
of particles as compared to the fastest approximations to Ewald summation, such as
Particle-mesh Ewald, which scale as $\mathcal O(N \log N)$~\cite{Darden1993,
Deserno1998}. However, it is well known that truncation of long-ranged Coulomb interactions in 
simulations can lead to severe artifacts~\cite{Neumann1980, 
Schreiber1992,Feller1996, Rodgers2008a, Spohr1997, VanDerSpoel2006,
Cisneros2014, Muscatello2011}. In particular, short-ranged methods often fail
for heterogeneous systems containing interfaces, even though they are known to
perform well in bulk equilibrium simulations provided that the parameters are
chosen carefully~\cite{Cisneros2014,Rodgers2008a,Muscatello2011,
Mendoza2008}. In simulations of the liquid--vapour interface, for
example, the Wolf method was found inadequate for predicting the electrostatic
potential and dipole orientations, regardless of the
choice of parameters~\cite{Takahashi2011}. In the context of local molecular
field (LMF) theory it has been demonstrated recently that averaged long-range effects can be taken into account self-consistently through an external potential~\cite{Rodgers2008a,Chen2006a}. 
In this approach, short-ranged interactions are modelled through a pairwise potential which bears 
strong similarities to the Wolf method~\cite{Rodgers2008}.
However, in the absence of the external potential the short-ranged method
failed to reproduce the correct results as obtained with Ewald
summation and molecules were found to overorient~\cite{Rodgers2008a}.

Here, using a full treatment of electrostatic interactions with Ewald summation
we investigate the validity of the electric fields and induced orientations
observed by Bresme and co-workers~\cite{Bresme2008,Muscatello2011a, Armstrong2013, Armstrong2013a,
Armstrong2015}.
The field
calculation requires especially careful consideration, as the large body of work
published thus far relies on the formulation which is inconsistent with the dynamics of the simulation~\cite{Bresme2008,Muscatello2011a, Armstrong2013, Armstrong2013a, Armstrong2015}. The correct
calculation of the field requires a modified kernel (rather than $r^{-1}$) that
is consistent with the effective truncated Coulomb
interactions~\cite{DeLeeuw1980, Neumann1984}. We discuss this issue in detail and carry out a comparison 
of the thermally induced fields and multipole moments as
obtained both with Ewald summation and the Wolf method.

Another important aspect that deserves consideration, is
the spatial averaging of the potential and the field. In order to resolve the
spatial variation of these quantities, it is advantageous to consider a quasi
one-dimensional setup to enhance sampling. Usually, the
charge density is first spatially averaged over small slabs (bins) and
then convoluted with an appropriate kernel to obtain, for example, the
potential~\cite{Wick2006, Wilson1988, Yeh1999,Glosli1996}.
As a consequence, the potential calculated in this way does, in
general, not represent the exact average over the individual bin. 
However, as we demonstrate in this work, calculating the exact analytical
average can be done straightforwardly for both summation methods and can lead to
huge reductions in the error bars for low resolutions. Therefore, this
approach frees us from the constraint of employing an unnecessarily high,
submolecular resolution.
 
The remainder of this paper is structured as follows: In
Sec.~\ref{sec:electrostatics}, we briefly summarise the electrostatic kernels
for Ewald summation and the Wolf method, respectively, and discuss important
differences using a simple model system.
Then, in Sec.~\ref{sec:spatavg}, we reduce the three-dimensional problem to one
spatial dimension employing symmetry properties of the setup. 
The two different multipole expansions considered in
this work are derived in Sec.~\ref{sec:multi}. The simulation protocol is
explained in Sec.~\ref{sec:sim} and all simulation results are presented in
Sec.~\ref{sec:res}.

\section{\label{sec:electrostatics}Electrostatic interactions}
In MD simulations, periodic boundary conditions (PBCs) are usually employed to
reduce finite-size or surface effects~\cite{Frenkel2002}. This implies that the
simulated system is infinite, but can be fully described with knowledge of the
state of a reference box. The electrostatic potential, $\Phi$, is governed by Poisson's equation,
\begin{equation}
\label{eq:poisson}
\nabla^2 \Phi = - 4\pi \rho_q,
\end{equation}
where $\rho_q$ is the charge density and all quantities are
expressed in Gaussian units. One way of determining the potential is to solve
this equation directly for the fictitious infinite system.
Alternatively, the task can be mapped onto the problem of finding a generalised
kernel or Green's function, $G$, compatible with a finite volume
with PBCs, considering nearest images only~\cite{Hummer1999b}.
Once $G$ is known, the potential and the field can then be calculated
as
\begin{equation}
\label{eq:phi}
\Phi(\mathbold r) = \int_\Omega \mathrm d^3r'\ G(\mathbold r - \mathbold r')
\rho_q(\mathbold r'),
\end{equation}
and 
\begin{equation}
\label{eq:E}
\mathbold E(\mathbold r) = - \nabla \Phi(\mathbold r),
\end{equation}
where $\Omega$ is the simulation box of volume $V$. Throughout
this work, we assume that PBCs are explicitly taken into account whenever
expressions that depend on an argument of the form $\mathbold r - \mathbold r'$
are evaluated~(see for example
Appendix~\ref{secapp:electrostatics}).

Although both approaches lead to the same result, there is an important conceptual difference:
In the former case, we consider the infinite system of charges
interacting with the potential that scales as $r^{-1}$ (in three dimensions)
plus surface term, whereas in the latter case, we only consider the charge distribution in our reference box with
an effective interaction. The periodicity of the setup is then
fully mimicked by the Green's function, which no longer decays as $r^{-1}$ and
is not even spherically symmetric.

Let us consider a charge-neutral
system consisting of $N$ molecules each
comprising $n$ partial charges $q_{i\alpha}$ located at positions $\mathbold
r_{i\alpha}$ ($i$ labels molecules and
$\alpha$ sites within a molecule). 
The total electrostatic energy is then given
by~\cite{Hummer1998,Hummer1999b}
\begin{alignat}{2}
U(\mathbold R) 
&=&& \phantom{{}+{}}  \frac 12 \sum_{i \neq j}\sum_{\alpha,\beta}
q_{i\alpha} q_{j\beta} \ G(\mathbold r_{i\alpha j\beta})\label{eq:UfromG} \\
& &&+\frac 12 \sum_{j}
\sum_{\alpha \neq \beta} q_{j\alpha} q_{j\beta} \left[ G(\mathbold
r_{j\alpha j\beta}) - \frac{1}{r_{j\alpha j\beta}} \right] \nonumber \\
&  &&+\frac 12 \sum_{j}\sum_{\alpha} q_{j
                         \alpha}^2 \lim_{r \to 0} \left[ G(\mathbold r) -
                         \frac{1}{r} \right], \nonumber
\end{alignat}
where $\mathbold r_{i\alpha j\beta} = \mathbold r_{j\beta} - \mathbold
r_{i\alpha}$ is the distance vector between the nearest pair of
images, $r = |\mathbold r|$ and $\mathbold R = (\mathbold r_{11}, \ldots,
\mathbold r_{Nn})$ is a $3n\times N$-dimensional vector. In the above equation we have
 omitted the summation bounds for readibility.

In Eq.~(\ref{eq:UfromG}) the surface term of de Leeuw and
co-workers~\cite{DeLeeuw1980} has been omitted, because we employ
conducting (tin-foil) boundary conditions.
We can see that the functional form of $G$ directly affects the forces, which
are calculated from the negative gradient of the energy, and therefore the
dynamics of the simulation.
In what follows, we briefly summarise the kernels for Ewald summation
and the Wolf method.

\subsection{\label{subsec:GE}Ewald summation}
Ewald summation is a numerical approximation to the exact solution of
Eq.~(\ref{eq:poisson}) for PBCs, whose Green's function is formally given by
\begin{alignat}{2}
\label{eq:Gpbc3D}
G_\text{PBC}(\mathbold r) = \frac{1}{V} \sum_{\mathbold k \neq \mathbf 0}
\frac{4\pi}{k^2} \text{e}^{i \mathbold k \cdot \mathbold r}.
\end{alignat}
Here, the summation extends over reciprocal vectors $\mathbold
k$ with components $k_\alpha = 2\pi p_\alpha/L_\alpha$, where
$p_\alpha$ is an integer and $L_\alpha$ the box size in
direction $\alpha$.
Introducing the convergence factor $\text{e}^{-k^2/4\eta^2}$, the expression is
split up into two terms, one of which is converted back to real space. This leads to the
representation~\cite{Hummer1998}
\begin{alignat}{2}
  G_\text{E,full}(\mathbold r) &= &&  \phantom{{}+{}} \sum_{\mathbold n } \frac{\text{erfc}(\eta
  | \mathbold r + \mathbold n|)}{| \mathbold r + \mathbold n|}  - \frac{\pi}{\eta^2 V}  
  \label{eq:Ga} \\
               & && + \frac {1}{V} \sum_{\mathbold k \neq \mathbf 0} \frac {4
               \pi}{k^2}  \text{e}^{-\frac{k^2}{4 \eta^2}}
                                                  \text{e}^{i \mathbold k \cdot
                                                  \mathbold r}, \nonumber
\end{alignat}
where $\mathbold n$ is a shift vector between a molecule and its periodic image and the summation
runs over all periodic images. Choosing $\eta$ carefully, it is possible to
achieve fast convergence of the first sum and small contributions for $\mathbold
n \neq \mathbf 0$.
If we ignore these terms and introduce a spherical cutoff, $r_\text{c}$, for
better performance, Eq.~(\ref{eq:Ga}) finally reduces to
\begin{alignat}{2}
  G_\text{E}(\mathbold r) &=&& \phantom{{}+{}} \Uptheta(r_\text{c} -r)
  \frac{\text{erfc}(\eta r )}{r} -
  \frac{\pi}{\eta^2 V} \label{eq:GEsim} \\ 
  & && + \frac {1}{V} \sum\limits_{\mathbold k \neq
  \mathbf 0} \frac {4 \pi}{k^2}  \text{e}^{-\frac{k^2}{4 \eta^2}} \text{e}^{i
  \mathbold k \cdot \mathbold r} \nonumber,
\end{alignat}
where $\Uptheta(r)$ is the Heaviside function.
Inserting this expression back into Eq.~(\ref{eq:UfromG}) yields the standard
Ewald summation expression~\cite{Hummer1999b} as presented in textbooks, e.g.~in
Ref.~\citenum{Frenkel2002}.

\subsection{\label{subsec:GW}Wolf method}
Wolf and co-workers showed that in a condensed ionic system the net Coulomb potential 
is effectively short-ranged~\cite{Wolf1999}. Based on this insight, they devised
a summation method that avoids the expensive \textit{k}-space term in Eq.~(\ref{eq:GEsim})
altogether. Instead, the potential is damped and shifted in a way that
enforces charge neutrality within the cutoff sphere for improved
convergence properties. The corresponding kernel is given by
\begin{align}
\label{eq:GW}
  G_\text{W}(\mathbold r) = 
  \Uptheta(r_\text{c}-r) \left[\frac{\text{erfc}(\zeta r )}{r}  - 
  \frac{\text{erfc}(\zeta r_\text{c})}{r_\text{c}} \right]
  \end{align}
and reproduces the correct Madelung energy as suggested by Wolf
and co-workers~\cite{Wolf1999}. Later the method was extended to
eliminate also higher-order multipoles inside the cutoff
sphere~\cite{Fukuda2011,Fukuda2013}.
However, it was pointed out that the entire approach embodies certain
assumptions on the underlying physical system~\cite{Fukuda2013}, such as 
the availability of charges outside the cutoff region for
screening~\cite{Elvira2014a}. Whether these assumptions are reasonable is not
always clear a priori, especially for inhomogeneuous systems such as the one
considered in this work.

We note that the first term in $G_\text{W}$ is identical to the one in $G_\text{E}$, 
although the optimal choice of the damping parameter, $\zeta$, is not
necessarily the same as for Ewald summation. A good value can be found by
analysing the convergence of the Madelung energy per
ion~\cite{Wolf1999}.
Furthermore, in the Wolf method the force is not exactly given by the negative
gradient of the potential energy. The reason for this inconsistency is that the expression
 $G^{'}_\text{W}(r)-G^{'}_\text{W}(r)|_{r=r_\text{c}}$ is used for the
 evaluation rather than $G^{'}_\text{W}(r)$ in order for the force to vanish at
 the cutoff distance~\cite{Wolf1999}.
 There are extensions of the Wolf method which address this
 issue~(for example Ref.~\onlinecite{Zahn2002}). However, given a reasonable
 combination of damping parameter and cutoff value, we expect the effects of
 this inconsistency on the electric field to be negligible.

\subsection{\label{subsec:model}Model system}
To illustrate the difference between the electrostatic kernels, we
consider a test case based on calculating the potential generated by a single SPC/E
water~\cite{Berendsen1987} molecule. This simple
example should draw attention to the fact that, for
an identical arrangement of charges, the results for the Wolf method sensitively
depend on the choice of kernel, damping parameter and cutoff radius.
The quality of the Wolf approximation to the electrostatic potential,
computed according to Eq.~(\ref{eq:phi}), is assessed by comparison with the
results of Ewald summation, which approximates the exact solution. 
 
Considering only a single molecule may seem atypical for the Wolf
method, since it relies on the idea that long-range contributions average out in
a dense system. However, this comparison serves as a guideline for the 
choice of new parameters which help us to reduce the dependence on this
crucial assumption. This is achieved by tuning the potential to get
better agreement with Ewald summation already on the level of a single molecule.
The comparison in Sec.~\ref{sec:res} will then allow us to assess the performance
of the Wolf method for a wider range of parameters, but it is not the intention
of this work to single out an optimal choice.

\begin{figure}[!t]
  \centering
   \includegraphics{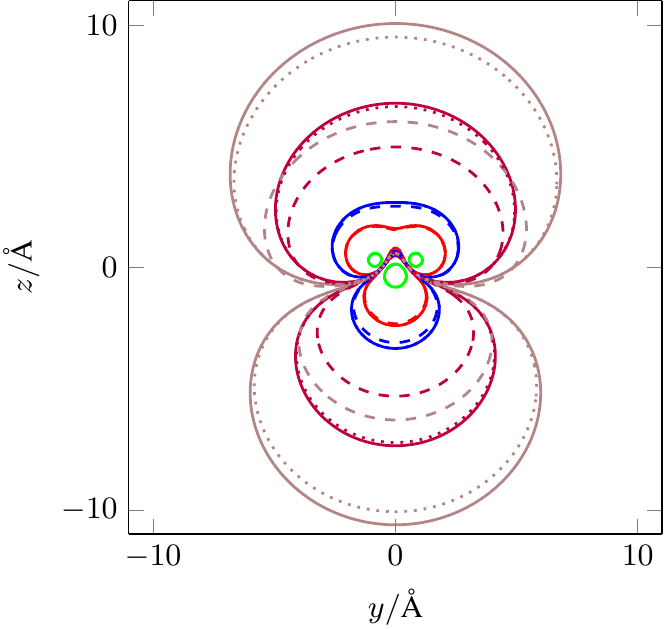}
   \caption{Absolute value of the electrostatic potential of a single SPC/E
   water molecule computed by Ewald summation (solid lines) and the Wolf method
   with $\zeta L = 1.0$ (dotted lines) and $\zeta L = 7.2$ (dashed
   lines). With increasing distance from the origin, the isolines of the
   potential correspond to the values $14.4~\text{V}$, $1.44~\text{V}$,
   $0.72~\text{V}$, $0.144~\text{V}$ and $0.072~\text{V}$,
   respectively.}
  \label{fig:PhiSpceModel}
\end{figure}

Figure~\ref{fig:PhiSpceModel} shows the potential due to a single
SPC/E water molecule in a fully periodic system.
The molecule is located at the centre of a rectangular simulation box with
dimensions $L= L_x = L_y = L_z/3 = 36.35~\text{\AA}$. The three charges, 
${q_\text{O} = 0.8476 q_\text{e}}$ and ${q_{\text{H}_{1/2}} =
-q_\text{O}/2}$, where $q_\text{e}$ is the elementary charge, are located in the ${x=0}$--plane at positions ${\mathbold
r_\text{O} = (0,0, -0.289)~\text{\AA}}$ and ${\mathbold r_{\text{H}_{1/2}}=
(0, \pm0.816, 0.289)~\text{\AA}}$, respectively.
Ewald summation was carried out taking
${r_\text{c}=L/2}$ with ${\eta L=5.85}$, and choosing the set of $\mathbold k$-vectors for Eq.~(\ref{eq:GEsim}) such that
the estimated relative error of the force was approximately $10^{-5}$. For the Wolf method, 
we compare two sets of parameters: $(\zeta L = 1.0, r_\text{c}=L/2)$ and $(\zeta
L = 7.2, r_\text{c}=11~\text{\AA})$. The latter combination was employed by
Armstrong and Bresme~\cite{Armstrong2013} and the former with considerably 
weaker damping and a larger cutoff is added for comparison.
We note that we also investigated the effects of a large cutoff combined with
strong damping, i.e. $(\zeta L = 7.2, r_\text{c}=L/2)$. However, we did not observe any substantial 
differences for the main results of this work compared
with the $11~\text{\AA}$ cutoff and therefore omitted the comparison.

It is obvious that for the strong damping (dashed lines) the potential decays
too quickly as compared to the result we get with Ewald summation (solid lines).
Only the short-range behaviour in the immediate vicinity of the molecule is
captured correctly. The weaker damping parameter (dotted lines), on the other
hand, yields a reasonable agreement with Ewald summation 
within a distance of about 6~\text{\AA} from the origin, but shows some
deviation further away. Employing even lower values for
$\zeta$, for example $0.5/L$, reduces the discrepancy between the outermost
contour lines only minimally (not shown). Since the value of the potential represented by the lowest contour
level in Fig.~\ref{fig:PhiSpceModel} corresponds to only $0.5\%$ of the highest one,
we conclude that the parameters $(\zeta L = 1.0, r_\text{c}=L/2)$ yield a
reasonable approximation to the Ewald result within the cutoff sphere
of $11~\text{\AA}$. Validation of both sets of parameters in bulk
simulations also reveals good agreement with Ewald summation (see
Appendix~\ref{secapp:valid}).

\section{\label{sec:spatavg}Spatial averaging}
\begin{figure}
  \centering
  \includegraphics{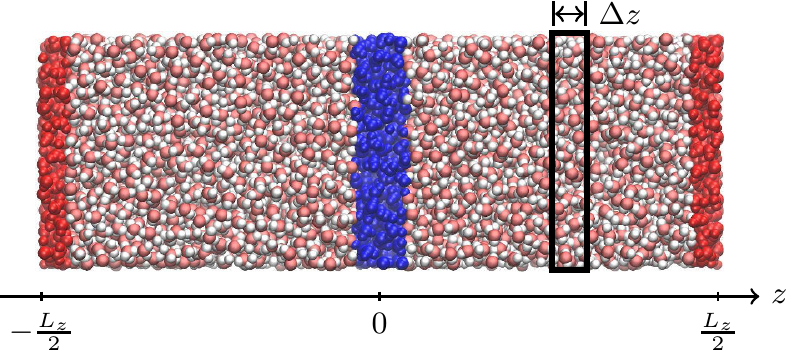}
  \caption{Simulation setup with a hot reservoir (coloured in red)
  wrapped around the boundaries and a cold reservoir (coloured in blue) in the centre of
  the simulation box. The superimposed rectangle (black solid
  lines) schematically illustrates a bin of width $\upDelta z$.}
\label{fig:box}
\end{figure}

Once the method to treat electrostatic interactions is chosen and
optimised, one typically wishes to improve the statistics of the collected
averages. For this purpose a simulation setup with high spatial symmetry is
advantageous~\cite{Armstrong2013}. In this work, we focus on the case where
the underlying three-dimensional problem can be reduced to one spatial
dimension, as illustrated in Fig.~\ref{fig:box}.
For such a system, the average charge density can only depend on $z$ for
sufficiently long simulation times, because the system is isotropic in all other directions. 
Therefore, this approach is justified only if one considers sufficiently long
simulations.
Assuming $\rho_q(\mathbold r') \equiv \rho_q(z')$, we can then rewrite Eq.~(\ref{eq:phi}) as
\begin{align}
\label{eq:phi3Dto1D}
\Phi (z) = \int\limits_{-L_z/2}^{L_z/2} \mathrm dz'\  {G_\text{1D}}(z-z') 
\rho_q(z'),
\end{align}
where we introduced the one-dimensional kernel
\begin{align}
\label{eq:phi3DGavg}
{G_\text{1D}}(z) = \int\limits_{-L_x/2}^{L_x/2} \mathrm dx'
\int\limits_{-L_y/2}^{L_y/2} \mathrm dy'\  G(x-x', y-y', z).
\end{align}
Taking the negative gradient of Eq.~(\ref{eq:phi3Dto1D}) yields the
electrostatic field
\begin{align}
\label{eq:Ez3Dto1D}
E_z(z) = -\int\limits_{-L_z/2}^{L_z/2} \mathrm dz'\  {G^{'}_\text{1D}}(z-z') 
\rho_q(z'),
\end{align}
where $G_\text{1D}^{'}$ denotes the derivative of $G_\text{1D}$. 
The above integrals can be evaluated readily for Ewald summation and the Wolf
method (see Appendix~\ref{secapp:electrostatics}). The results can be improved considerably by averaging
the potential and the microscopic field over small spatial regions, such that we
obtain the macroscopic Maxwell field for the latter. The centre
of each control volume then represents its exact spatial average. To this end, we consider $N_\text{b}$ bins of width
$\upDelta z$, as depicted in Fig.~\ref{fig:box}.
The lower and upper boundaries of bin $j$, where $j=1,\ldots,N_\text{b}$, are given by $z_{j,1} =
-L_z/2 + (j-1) \upDelta z$ and $z_{j,2} = z_{j,1} +\upDelta z$, respectively. The spatial average of
the potential over bin $j$ is then given by
\begin{subequations}
\label{eq:phi_ex_avg_der}
\begin{align}
{\bar \Phi}_j  &=  \frac{1}{\upDelta z} \int\limits_{z_{j,1}}^{z_{j,2}} \mathrm
dz \ \Phi(z) \\
               &= \int\limits_{-L_z/2}^{L_z/2} \mathrm dz' \
               {{\bar G}_{\text{1D},j}}(z') \rho_q(z'), 
\end{align}
\end{subequations}
where the overbar denotes the spatially averaged kernel
\begin{align}
\label{eq:G1DAnaAvg}
{\bar G}_{\text{1D},j}(z') = & \frac{1}{\upDelta z} 
\int\limits_{z_{j,1}}^{z_{j,2}} \mathrm dz\ {G_\text{1D}}(z-z').
\end{align}
For our effectively one-dimensional system of point charges, we can decompose
the charge density according to
\begin{align}
\label{eq:rhoq_z}
\rho_q(z) = \frac{1}{L_x L_y} \sum_{i} q_i\ \delta(z-z_i),
\end{align}
where $\delta(z)$ is the one-dimensional Dirac delta function.
Inserting this expression back into our previous result for the potential yields
\begin{align}
\label{eq:phi_ex_avg}
 {\bar \Phi}_j &= \frac{1}{L_x L_y} \sum_{i}
 q_i \ {\bar G}_{\text{1D},j}(z_i) .
\end{align}
Analogously, the averaged field is given by
\begin{align}
\label{eq:Ez_ex_avg}
 {\bar E}_{z,j} &= -\frac{1}{L_x L_y} \sum_{i} q_i \ {\bar
 G}^{'}_{\text{1D},j}(z_i).
\end{align}
The corresponding expressions for $\bar G_\text{1D}$ and ${\bar
G}^{'}_\text{1D}$ for Ewald summation are derived in
Appendix~\ref{secapp:electrostatics}. The above averages for potential and field depend on all particle positions and therefore implicitly on time. The time average of any quantity $X$ is defined as
\begin{align}
\label{eq:Xtavg}
\langle { X}\rangle &= \frac{1}{\tau} \int\limits_{0}^\tau \mathrm dt \ { X}(t),
\end{align}
where $\tau$ is the total simulation time of the production run. It is
straightforward to evaluate $\langle {\bar \Phi}_{j} \rangle$ and $\langle {\bar
E}_{z,j} \rangle$ for the discrete trajectory obtained from the NEMD simulation.

\section{\label{sec:multi}Multipole expansion}
\begin{figure}
  \centering
  \includegraphics{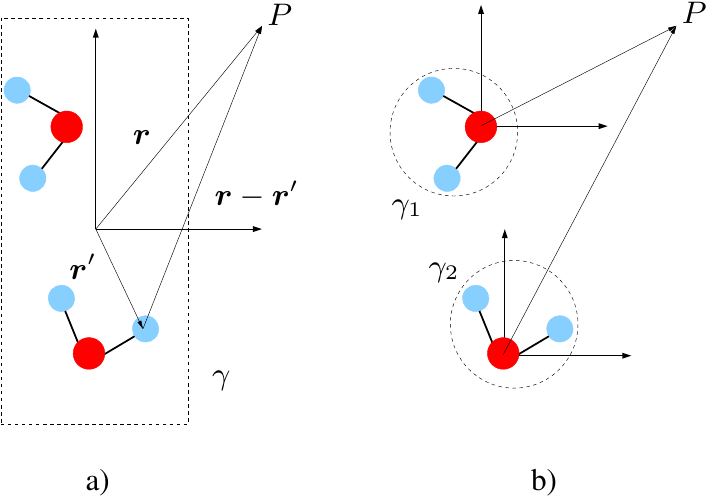}
  \caption{Illustration of two different multipole expansions: a)
	with respect to the centre of the region $\gamma$ (``slab expansion'') and
	b) for each molecule
	$\gamma_j$ individually with the oxygen site at the
	origin (``molecule expansion''). Both approaches give rise to
	the same field at a distant point $P$.}
  \label{fig:multi}
\end{figure}

In what follows, we outline how the exact potential, as calculated
from the charge density, can be decomposed into individual multipole
contributions. This helps us to gain insight into how the
alignment of the molecules with respect to the temperature gradient affects the field. 
We consider two different expansions for comparison which are illustrated in
Fig.~\ref{fig:multi}. In the slab expansion
(Fig.~\ref{fig:multi}a), the multipole moments due to the charges located inside a bin are calculated relative to its centre.
In the molecule expansion (Fig.~\ref{fig:multi}b),
separate multipole expansions are carried out for each individual molecule
and the multipoles are located at the respective oxygen sites. If all moments
were considered in the expansion, both approaches would give rise to the same
potential at a distant point $P$. We note that
both types of expansion have already been considered in the
past for interfacial systems~\cite{Wilson1989,Glosli1996}.
However, here we use a more general formulation~\cite{Smith1998} which is also
applicable to modified kernels representing truncated Coulomb interactions.

The potential generated by a charge distribution enclosed in a
volume $\gamma$ is given by
\begin{equation}
\label{eq:PhiMol}
\Phi(\mathbold r) =  \int_\gamma \mathrm d^3r'\ G(\mathbold r - \mathbold r')
\rho_q(\mathbold r').
\end{equation}
From this equation we can obtain the contributions of the individual
multipole moments by expanding $G(\mathbold r- \mathbold r')$ into a Taylor series around $\mathbold r$,
\begin{alignat}{2}
\Phi(\mathbold r) &\approx&& 
                \phantom{{}-{}}  G(\mathbold r) \underbrace{\int_\gamma \mathrm
                d^3r' \rho_q(\mathbold r')}_{q}  \label{eq:moments3D} \\
                & && - \sum_{\alpha} \nabla_\alpha
                G(\mathbold r)  \underbrace{\int_\gamma \mathrm d^3r' r_\alpha' 
                \rho_q(\mathbold r')}_{\mu_\alpha} \nonumber \\
                 & && + \sum_{\alpha,\beta}  
                \nabla_\alpha \nabla_\beta G(\mathbold
                r)  \underbrace{\frac 12 \int_\gamma \mathrm d^3r' r_\alpha'r_\beta'
                \rho_q(\mathbold r')}_{Q_{\alpha \beta}}, \nonumber
\end{alignat}
where $q$ is the total charge in $\gamma$, $\bm \mu$ the dipole moment and ${Q}$
the quadrupole moment.
The symbol $\nabla_\alpha$ denotes the derivative with respect to the
Cartesian component $r_\alpha$. Moving the origin of the charge distribution to
$\tilde {\mathbold r}$ and taking into account
the symmetry properties of our effectively one-dimensional system, we find
\begin{alignat}{2}
L_x L_y \Phi(z) &\approx&& \phantom{{}+{}} \underbrace{{G_\text{1D}}(z-\tilde
{z}) \ q}_{\text{monopole contribution}} - \underbrace{ G^{'}_\text{1D}(z-\tilde
{z})\ \mu_{z}}_{\text{dipole contribution}} \label{eq:PhiMulti} \\
& && + \underbrace{G^{''}_\text{1D}(z -\tilde {z})\ {
Q}_{zz}}_{\text{quadrupole contribution}} \nonumber.
\end{alignat}

From the simulated trajectory, we then compute time averages of the
multipole densities ${\bar \rho}_{q,j}$, ${\bar \rho}_{\mu,j}$ and ${\bar
\rho}_{Q,j}$ for the monopole, dipole and
 quadrupole moments of every bin $j$, respectively. 
Before defining these quantities, we first introduce some
 additional notation to distinguish between the two types of expansion. We use superscripts
$\cdot^{(m)}$, where $m=1$ for slabs (Fig.~\ref{fig:multi}a) and $m=2$ for molecules (Fig.~\ref{fig:multi}b). 
The density of $X= q,\mu_z, Q_{zz}$ [cf. Eq.~(\ref{eq:moments3D})] is then given by
\begin{equation}
{\bar \rho}_{X,j}^{(1)} = \frac{1}{\upDelta v} \times \left\{
\text{moment of bin $j$} \right\}
\end{equation}
for the case $m=1$ and 
\begin{equation}
{\bar \rho}_{X,j}^{(2)}  = \frac{1}{\upDelta v} \times \left\{
\text{sum of molecular moments in bin $j$} \right\}
\end{equation}
for the case $m=2$, where
$\upDelta v = L_x L_y \upDelta z$ is the volume of the bin. 
Since we only consider the multipole moments $q$, $\mu_z$ and $Q_{zz}$, 
from now on we omit the subscripts for readability.

In general, the multipole moments depend
on the way the charge distribution is
partitioned~\cite{Jackson1998,Spaldin2012} and consequently the multipole
densities for slabs and molecules are not directly comparable. For example, the quadrupole moment of a reference
bin will, in general, not be equal to the sum of the molecular
quadrupole moments. Furthermore, we make an intentional, small
mistake in the evaluation of ${\bar \rho}_{\mu,j}^{(2)}$ and ${\bar \rho}_{Q,j}^{(2)}$ for the sake of computational convenience, 
because we ignore the precise location of the molecular moments within the bin
$j$. However, as we will see in Sec.~\ref{sec:res}, the error in the
electrostatic potential introduced by this approximation is negligible.

The electrostatic potential (at the centre of bin $j$) is then calculated as the sum of the three
contributions in Eq.~(\ref{eq:PhiMulti}),
\begin{alignat}{2}
\label{eq:avgphimulti}
\Phi^{(m)}_j =  \Phi^{(m)}_{q,j} 
+ \Phi^{(m)}_{\mu, j} +  \Phi^{(m)}_{Q,j},
\end{alignat}
which are given by
\begin{subequations}
\label{eq:avgphimulticontr}
\begin{alignat}{2}
 \Phi^{(m)}_{q,j} &=&& \phantom{{}-{}} \upDelta z \sum_{l=1}^{N_\text{b}} {
G_\text{1D}}(z_j-z_l) {\bar \rho}_{q,l}^{(m)}, \\
\Phi^{(m)}_{\mu,j} &=&&  -\upDelta z \sum_{l=1}^{N_\text{b}} 
G_\text{1D}^{'}(z_j-z_l) {\bar \rho}_{\mu,l}^{(m)}, \\
 \Phi^{(m)}_{Q,j}
 &=&&  \phantom{{}-{}}  \upDelta z \sum_{l=1}^{N_\text{b}} G_\text{1D}^{''}(z_j-z_l)
 {\bar \rho}_{Q,l}^{(m)},
\end{alignat}
\end{subequations}
respectively.
Since the molecules are charge-neutral, it follows that all values $ \rho_{q,j}^{(2)}
$ and consequently $ \Phi_{q,j}^{(2)}$ vanish
identically.

\section{\label{sec:sim}Simulation protocol}
For production runs, we prepared the system in the same state
as Armstrong and Bresme~\cite{Armstrong2013} in order to carry out a quantitative comparison.
 The simulation box (Fig.~\ref{fig:box}) has exactly the same 
 dimensions as the one used for the model system. For two of the three NEMD
 simulations we used the Wolf method and the remaining one was performed with
 Ewald summation (the relevant parameters are summarised in Sec.~\ref{subsec:model}).
 Lennard-Jones interactions were truncated at
 $11~\text{\AA}$ in all cases.
 The box contains $N=4500$ SPC/E molecules resulting in a mass density of $\rho_m = 0.934~\text{g}/\text{cm}^3$.
 All simulations were carried out using a modified version of the software
 package LAMMPS (9Dec14)~\cite{Plimpton1995} which we augmented with the eHEX/a
 algorithm~\cite{Wirnsberger2015}.

\subsection{\label{sec:equi}Equilibration}
The system was first equilibrated and validated. Starting from
an initial lattice structure with zero linear momentum, we integrated the equations of motion with 
the velocity Verlet algorithm~\cite{Swope1982a} employing a timestep of
${\upDelta t=1~\text{fs}}$. For the first $20$~ps we rescaled the velocities to drive the
system close to the target temperature of $400~\text{K}$. This was followed by
a short 200~ps \textit{NpT} run using a Nos\'e--Hoover
thermostat with a relaxation time of $\tau_T =
1~\text{ps}$ and a Nos\'e--Hoover barostat with a relaxation
time of $\tau_p = 2.5~\text{ps}$~\cite{Nose1984, Hoover1985}. We then rescaled
the box to the target dimensions and carried out a $500$~ps \textit{NVT} run during which we monitored the average system energy.
Next, we adjusted the kinetic energy of the last configuration by velocity
rescaling and used it as input for another $1$~ns \textit{NVE} equilibration run. 
The average temperature during this run was $T = (400 \pm 0.1)~\text{K}$, where
the error bar was estimated using block average analysis~\cite{Frenkel2002}. 
We computed the pair-correlation function, the velocity autocorrelation
function, the dielectric constant and the distance-dependent Kirkwood $g$-factor (see Appendix~\ref{secapp:valid}). 
The validation suggests that our implementation is correct and our
choice of parameters reasonable.

\subsection{\label{subsec:thermo}Non-equilibrium stationary
state} 
To investigate the effect of a thermal gradient after the
equilibration, the system was driven to a non-equilibrium
stationary state by imposing a constant heat flux between two reservoirs, $\Gamma_1$ and $\Gamma_2$ (Fig.~\ref{fig:box}). This was achieved by
introducing an additional force, $\mathbold g_i$, to the equations of
motion~\cite{Wirnsberger2015}, such that
\begin{subequations}
\begin{align}
  \dot{\mathbold r}_i &= \mathbold v_i, \label{eq:hexvcont3a} \\
  \dot{\mathbold v}_i &= \frac{\mathbold f_i}{m_i} + \frac{\mathbold g_i}{m_i},
  \label{eq:hexvcont3b}
\end{align}
\end{subequations}
where $m_i$ is the mass of atom $i$ and $\mathbold f_i$ the force calculated
from the potential. The thermostatting force is defined as
\begin{equation}
  \label{eq:eta}
  \mathbold g_i = 
  \begin{cases} m_i   \frac{\mathcal F_{\Gamma_{k(\mathbold r_i)}}}{2 \mathcal
  K_{\Gamma_{k(\mathbold r_i)}}} \left(\mathbold v_i - \mathbold v_{\Gamma_{k(\mathbold
  r_i)}} \right) &\mbox{if $k(\mathbold r_i) > 0$,} \\
   0 &\mbox{otherwise, } 
  \end{cases}
\end{equation}
where $k(\mathbold r_i) \in \{0,1,2\}$ is an indicator function which maps the
particle to the region $\Gamma_k$ in which it is located and $\mathcal
F_{\Gamma_k}$ is a constant energy flux into $\Gamma_k$. Those
parts of the simulation box which are not thermostatted are labelled with
$\Gamma_0$.
The non-translational kinetic energy of the region $\Gamma_k$ is given by 
\begin{equation}
{\mathcal K}_{\Gamma_k} = \sum_{{i \in \gamma_k}} \frac{ m_i v_i^2}{2} - \frac{m_{\Gamma_k} v_{\Gamma_k}^2}{2}
\label{eq:Ekinnt},
\end{equation}
where the quantities $\mathbold v_{\Gamma_k}$ and $m_{\Gamma_k}$ are the centre
of mass velocity and the total mass of $\Gamma_k$, respectively, and the index
set $\gamma_k$ comprises all particles which are located inside that
region~\cite{Wirnsberger2015}.

The equations were solved numerically with our recently proposed eHEX/a
algorithm~\cite{Wirnsberger2015} with a timestep of $\upDelta t = 2~\text{fs}$.
For the symmetric setup shown in Fig.~\ref{fig:box}, the heat flux is trivially 
related to the energy flow into the reservoir by
\begin{equation}
J_{Q,z} = \frac{F_{\Gamma_1}} {2 L_x L_y},
 \end{equation}
where the factor of 2 in the denominator accounts for the periodic setup.  
After switching on the thermostat, we waited for 10~ns for any transient
behaviour to disappear before starting with the $\tau =60~\text{ns}$ production
run. The energy conservation was excellent
($|\upDelta E/E| \approx 0.005\%$) and the centre
of mass velocity of the simulation box remained close to machine precision throughout the simulation. 
The heat fluxes are input parameters of the eHEX algorithm which were adjusted
by trial and error. The employed values are summarised in
Tab.~\ref{tab:T}.
\begin{table}
\caption{\label{tab:T}Imposed heat fluxes and measured values for the tempature
gradients. We note that our heat flux for the Wolf ($\zeta L = 7.2$)
run is about 1.7\% larger than the value used by
Armstrong and Bresme~\cite{Armstrong2013}.}
\begin{ruledtabular}
\begin{tabular}{lcc}
  &$J_{Q,z}$ ($10^{10}~\text{W}/{\text{m}^2}$)& $\nabla T$ (K/\AA)  \\ \hline
Ewald 		           & 4.243 & $-5.14 \pm 0.04$  \\
Wolf ($\zeta L = 1.0$)  & 4.166 & $-5.17 \pm 0.04$  \\
Wolf ($\zeta L = 7.2$)  & 3.875 & $-5.18 \pm 0.04$  \\
\end{tabular}
\end{ruledtabular}
\end{table}
We note that lower heat fluxes are required for the Wolf method in order to
achieve the same temperature gradient as for Ewald summation. 
This is consistent with the observation that the truncation of 
electrostatic interactions results in lower thermal
conductivities~\cite{Muscatello2011}.

\section{\label{sec:res}Results}
In this section, we present the key results for the temperature and density
profiles (Sec.~\ref{subsec:restemp}), the multipole expansions
(Sec.~\ref{subsec:resmoments}), the potential (Sec.~\ref{subsec:respot}), the
field (Sec.~\ref{subsec:resfield}) and the polarisation
(Sec.~\ref{subsec:respol}). We estimated error bars for all
results in this section. To this end we divided the entire
trajectory into 600 blocks (of length $100~\text{ps}$) and assumed the
results for the individual blocks to be uncorrelated. The size
of the individual error bar then corresponds to twice the standard deviation of the mean. This estimate
comprises the statistical error as well as the methodological
error arising, for example, from the employed quadrature. 

\subsection{\label{subsec:restemp}Temperature and density}
\begin{figure*}
  \centering
    \includegraphics{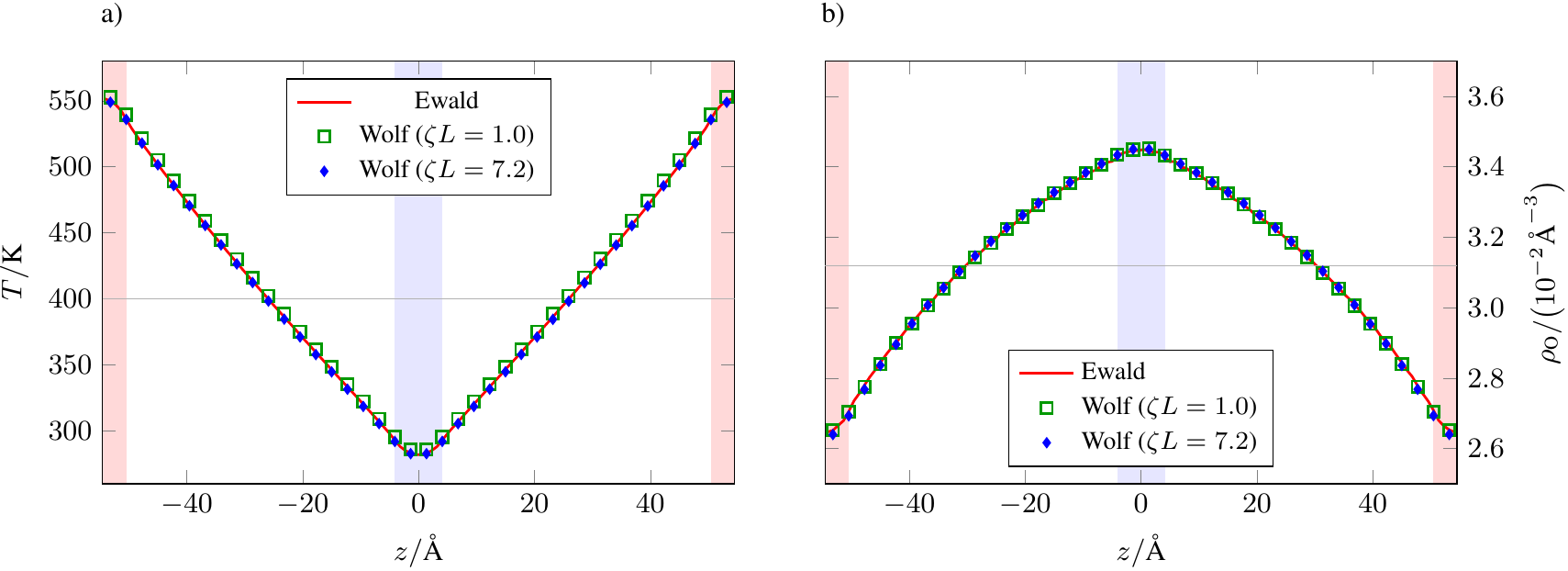}
   \caption{\label{fig:Trho}Spatial variation of a) temperature and b) oxygen
   number density obtained with Ewald summation and the Wolf method. The
   horizontal lines indicate the spatial and temporal equilibrium averages of
   the temperature and the number density, respectively.
   The vertical stripes indicate the locations of the hot
   (coloured in red) and cold (coloured in blue) reservoirs.}
\end{figure*}

Figures~\ref{fig:Trho}a-b show the spatial variations in
temperature and density along the $z$-direction with a resolution of $\upDelta z = 2.73~\text{\AA}$
($N_\text{b} = 40$).
The temperature of an individual bin was calculated from the non-translational 
kinetic energy of the atoms inside that bin~\cite{Wirnsberger2015}. There are
only small differences between the results obtained with the Ewald and Wolf methods. The peak temperature
at the centre of the hot reservoir is about $552~\text{K}$ and the lowest temperature at the centre of
the cold reservoir is about $285~\text{K}$ (Fig.~\ref{fig:Trho}a).
The temperature profile is linear outside the reservoirs and symmetric with
respect to the origin of the simulation box, which is in accordance with the setup.  

The measured average number densities (Fig.~\ref{fig:Trho}b) obtained with Ewald
summation and the Wolf method agree well apart from slight differences in the
vicinity of the cold reservoir. The mass density varies by up to 15\% (cold
reservoir) with respect to $\rho_m$. We note that on this scale, we did not
observe any appreciable discontinuities of the temperature or density close to the reservoir boundaries, although the
thermostatting force is discontinuous.

\subsection{\label{subsec:resmoments}Molecular orientation and multipole
moments}
\begin{figure*}
   \centering
   \includegraphics{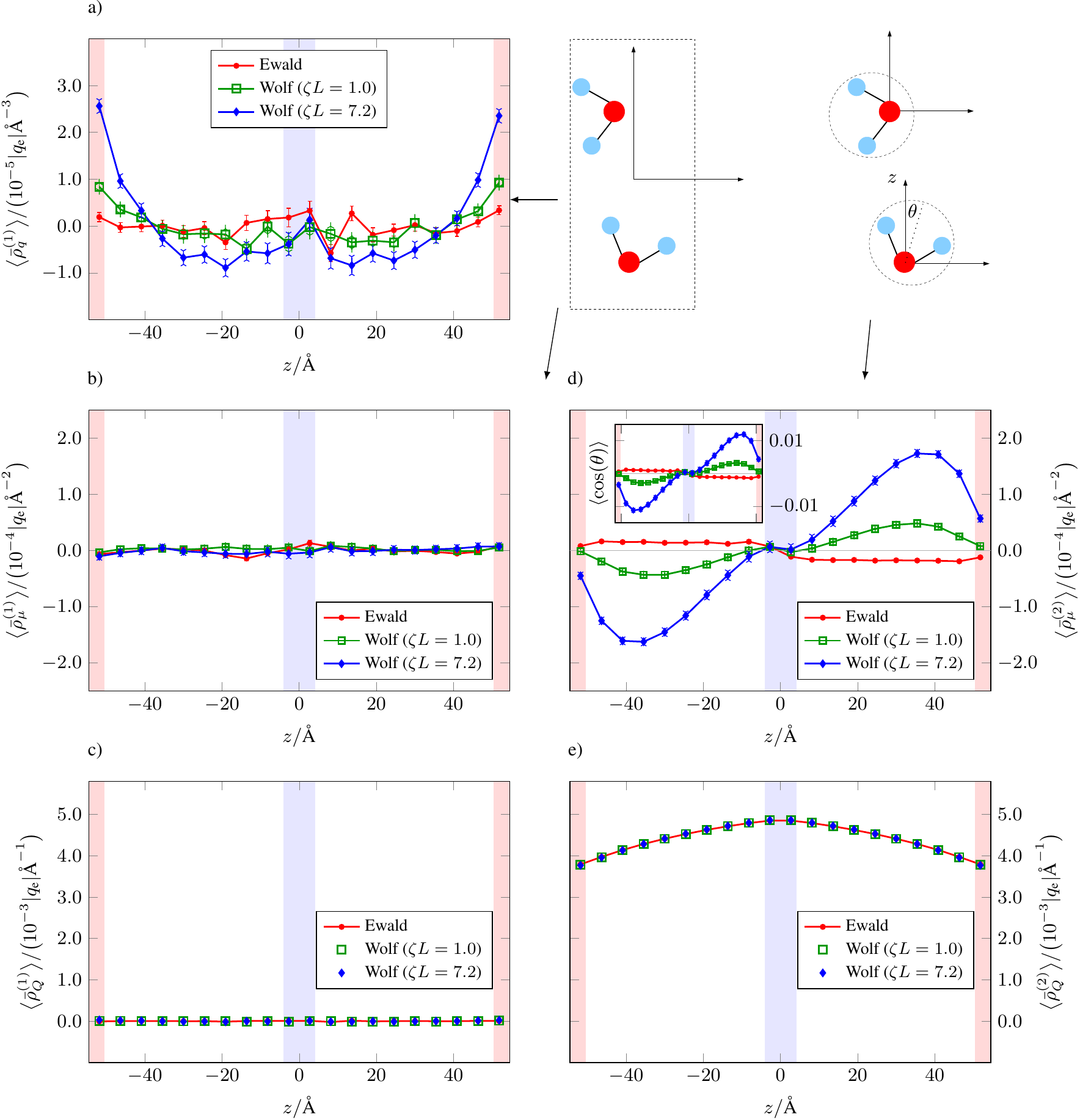}
     \caption{\label{fig:multicomb}Multipole densities for the slab expansion
     (left column) and molecule expansion (right column). The
     panels a-c) show the charge density, dipole density and quadrupole
     density, respectively, for the slab expansion.
     The panels d) and e) show the dipole density and quadrupole density, respectively, for the
     molecule expansion.
     The inset in panel d) shows the average molecular alignment with the temperature gradient. Horizontal lines in the inset and the full
     figure were added to highlight the symmetry of the result.}
 \end{figure*}

In this section, we discuss the induced molecular alignment and multipole moments due to the
thermal gradient for both expansions in Fig.~\ref{fig:multi}. The left column in
Fig.~\ref{fig:multicomb} corresponds to the slab (centre-of-bin) expansion and
the right column to the molecule expansion. The monopole in the
molecule expansion vanishes identically, hence it is not shown. The spatial
variations of all quantities are shown with a resolution of $\upDelta z =5.45~\text{\AA}$ ($N_\text{b} = 20$).

Let us consider the time 
averaged charge density for slabs first (Fig.~\ref{fig:multicomb}a).
For Ewald summation the error of the average is 
so large that it swamps the signal even after 60~ns of
simulation time. We also note that the curve is not symmetric in the
vicinity of the cold reservoir within the statistical uncertainty shown in the
plot. We believe that this may be due to the fact that we computed the error
bars as if neighbouring bins were independent, which is not the case, because
molecules are charge neutral. The real error bars may be larger due to long-wavelength fluctuations. 
We confirmed that the results become symmetric (within the
statistical error) upon doubling the simulation time. 

For the Wolf
method there is an accumulation of positive charge in the vicinity of the hot reservoir, which is enhanced by stronger damping. This result agrees qualitatively with the findings of Rodgers 
and Weeks for a different inhomogeneous system, where the authors compared the (Gaussian-smoothed) charge density obtained with Gaussian-truncated (GT) water 
to that of Ewald summation~\cite{Rodgers2008a}. Furthermore, we note that the error bar increases by about one
order of magnitude upon refining the resolution by a factor of 10, which corresponds to $\upDelta z \approx 0.54~\text{\AA}$ ($N_\text{b} = 200$) used by Armstrong and Bresme~\cite{Armstrong2013}.

Figures~\ref{fig:multicomb}b and d show the dipole densities for both
expansions, respectively. For the slabs (Fig.~\ref{fig:multicomb}b), there is no noticable
trend of the dipole density within the statistical uncertainity.
However, for the molecule expansion (Fig.~\ref{fig:multicomb}d) we find a strong
disagreement between the two electrostatic kernels.
For this case, we also quantified the average molecular alignment
using the order parameter~\cite{Romer2012}
\begin{equation}
\label{eq:molal}
\cos(\theta) = \mathbold n \cdot \mathbold e_z,
\end{equation}
where $\mathbold n = \mathbold \mu/\mu$ defines the orientation of a molecule
and $\mathbold e_z$ is the unit vector in the direction of the temperature
gradient. In the case of Ewald
summation molecules, on average, point to the cold reservoir and the alignment 
is fairly constant outside the reservoirs (see inset in 
Fig.~\ref{fig:multicomb}d).
The Wolf method entirely fails to capture this behaviour. 
For the wide range of parameters considered in this work (including the ones previously employed in the literature), 
the method predicts opposite orientations and overestimates the magnitude of alignment by a factor of about 7 
for the strong damping. Employing a lower value for the damping parameter
reduces the overestimation, but cannot correct the wrong sign. We also note that
our results for the average molecular orientation (inset in 
Fig.~\ref{fig:multicomb}d) are in agreement with
the ones reported by Armstrong and Bresme~\cite{Armstrong2013}.

The quadrupole densities, shown in Figs~\ref{fig:multicomb}c and e,
 agree well with each other within each expansion. Similarily to the dipole
 density, considering slabs for the expansion (Fig.~\ref{fig:multicomb}c) yields results which are
negligible compared to the molecule expansion (Fig.~\ref{fig:multicomb}e). We
note that in the latter case, the profile is proportional to the
oxygen number density (Fig.~\ref{fig:Trho}b) and can lead to considerable
contributions to the potential.

Repeating our simulation
with Ewald summation and vacuum boundary conditions (see Refs~\onlinecite{Hummer1999b, DeLeeuw1980} for more
details), we found consistent results for the multipole densities (not
shown). We can therefore rule out any artifacts arising from the boundary
conditions at infinity on the results shown in this section. However, we
noticed that the statistical error of the molecular dipole density decays much
faster for vacuum boundary conditions relative to tin-foil boundary conditions.

\subsection{\label{subsec:respot}Electrostatic potential}
In the previous section, we analysed the thermally 
induced multipole moments for two different multipole expansions, namely slabs
and molecules. 
The aim of this section is to compare three different ways of
calculating the electrostatic potential: Firstly, we consider the
exact analytical average given by Eq.~(\ref{eq:phi_ex_avg}). Secondly, we
approximate the potential using only the average charge density given by the
slab expansion, Eq.~(\ref{eq:avgphimulticontr}a), 
which is the approach regularly employed in the literature~\cite{Wick2006,
Wilson1988,Yeh1999, Armstrong2013}. Thirdly, we approximate the potential using also the dipole
and quadrupole densities, i.e. Eqs~(\ref{eq:avgphimulticontr}b-c).

\begin{figure*} 
  \centering
    \includegraphics{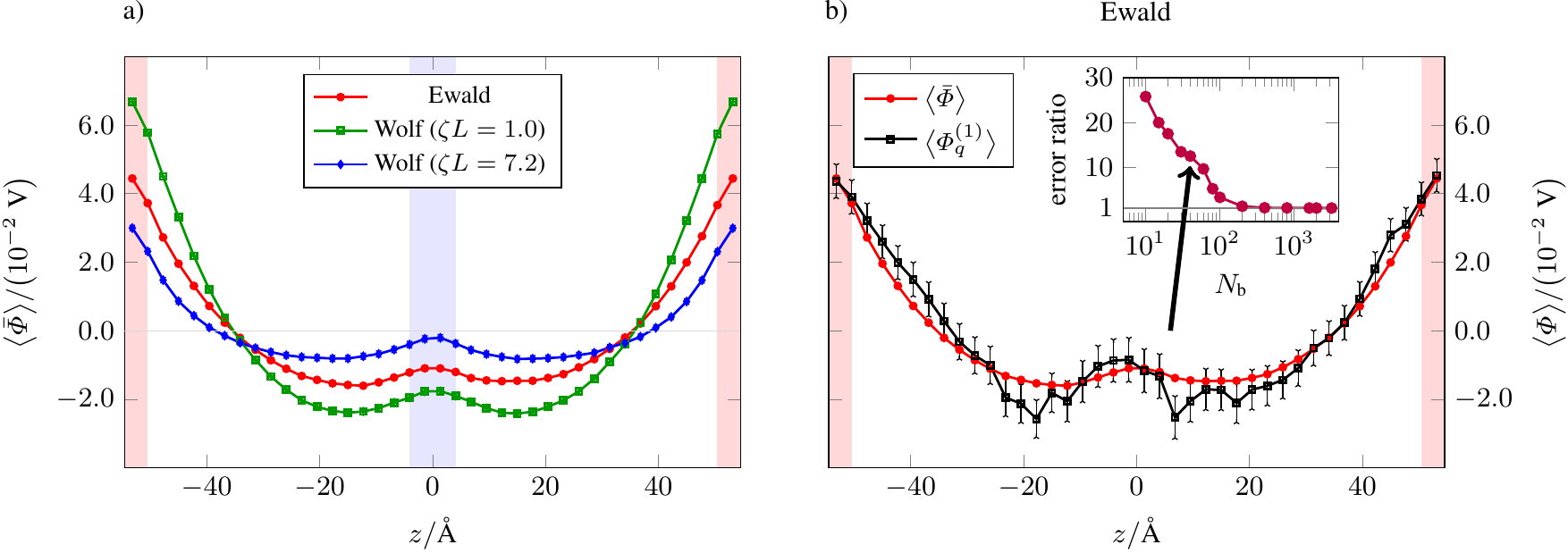}
   \caption{\label{fig:phiex}The exact potential is shown in panel a) and a
   comparison between the potential calculated solely from the monopole
   density in the slab expansion and the exact result calculated with Ewald
   summation is shown in b). The inset compares the ratio of the
   maximum errors which were calculated from 600 blocks of length $100~\text{ps}$ as a function of
   the number of bins. The arrow indicates the error ratio for the resolution
   shown in the full figure.}
\end{figure*}

Let us consider the results for the exact calculation first, which are shown in
Fig.~\ref{fig:phiex}a. All graphs are symmetric with respect to the origin of the
simulation box and periodic, indicating that the field vanishes at the
centres of the reservoirs. Although the shape of the potential predicted by the short-ranged
method is similar to the one for Ewald summation, the results are
sensitive to the choice of damping parameter. Weak damping
overestimates the potential, whereas strong damping leads to an underestimation. Both our 
choices fail to reproduce the Ewald summation result correctly, although it
seems plausible that intermediate values for the damping
parameter could lead to a better agreement.

Figure~\ref{fig:phiex}b compares (for Ewald summation) the exact result for the
electrostatic potential to that given by the monopole density in
the slab expansion.
We recall that the latter approach corresponds to averaging the
charge density first and integrating it with the appropriate kernel afterwards
[Eq.~(\ref{eq:avgphimulticontr})a].
It is clear from comparison of the two curves including error bars
that the exact calculation yields a huge improvement over the approximation. For the
resolution shown in the plot ($N_\text{b}=40$, $\upDelta z=2.73~\text{\AA}$), the error bars are reduced by
more than one order of magnitude. The inset shows the ratio of the maximum error of the
approximation to the maximum error of the exact calculation as a function of
the number of bins.
(We define the maximum error to be half the length of the largest error bar throughout the entire interval.) For a very
low resolution of 10 grid points ($\upDelta z=10.9~\text{\AA}$), 
the maximum error decreases by about a factor of 26. For high
resolutions of $\upDelta z \leq 0.5~\text{\AA}$ the error ratio approaches
unity implying that both methods become comparable, which is the expected
behaviour in the limit $\upDelta z \to 0$. At the same time the magnitude of
the error naturally increases for higher resolutions because fewer molecules contribute to a particular bin (for 400 bins the maximum error
increases by about $50\%$ as compared to the resolution of 40 bins shown in
the figure).

\begin{figure*}
  \centering
    \includegraphics{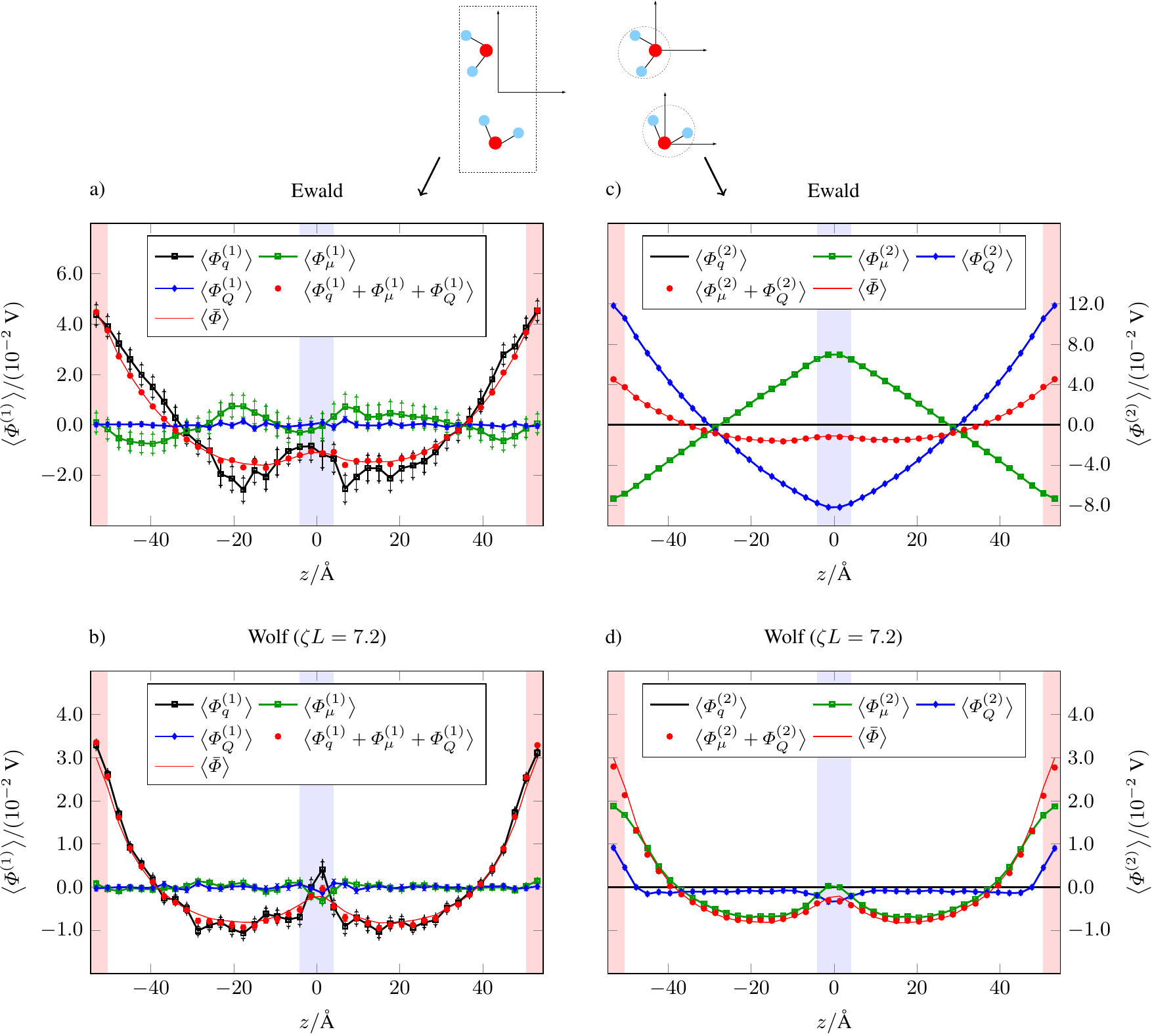}
   \caption{\label{fig:phimulticomb}Individual contributions to
   the potential for the slab expansion (left column) and molecule expansion (right column).
   The results for Ewald summation are shown in panels a) and c) in the first
   row and for the Wolf method in panels b) and d) in the second row.}
\end{figure*}

Given that molecules point, on average, in opposite directions
for the two electrostatic kernels (Fig.~\ref{fig:multicomb}d), it is
counterintuitive that the potentials are qualitatively comparable.
To understand the origin of this seeming contradiction, we singled
out the individual multipole contributions, which are illustrated in
Figs~\ref{fig:phimulticomb}a-d for both expansions. Let us consider the slab
expansion first. For both electrostatic kernels
(Figs~\ref{fig:phimulticomb}a-b) we found the monopole contribution (black
curve) to capture the exact potential (red line) reasonably well for the chosen
spatial resolution ($N_\text{b}=40$, $\upDelta z=2.73~\text{\AA}$). However, if we consider 
a point dipole and a point quadrupole (representative for the respective bin
average) in addition to the point monopole located at the centre of each bin, we
obtain a much better approximation to the exact result (red circles). In fact,
for Ewald summation we recover the exact potential almost perfectly, whereas
we observe an overshoot inside the hot reservoir for the Wolf
method.
We believe that a more accurate approximation for the
short-ranged method might be obtained by considering octupole and hexadecapole
contributions in addition, but we did not investigate this further.

The situation changes entirely for the molecule expansion shown in
Figs~\ref{fig:phimulticomb}c-d, where the monopole contribution is zero. For
Ewald summation (Fig.~\ref{fig:phimulticomb}c), the dipole density leads to a linear potential
outside the reservoirs (green curve) corresponding to a negative field in the
left half of the simulation box. However, close to the hot reservoir the
quadrupole contribution (blue curve) outweighs the dipole contribution causing
the slope of the overall potential to be negative and therefore
the field to be positive. In the vicinity of the cold reservoir the dipole
contribution dominates over the quadrupole contribution and the field is negative.
The sum of both terms (red circles) agrees perfectly with the exact
average (red line). For the Wolf method we found that the quadrupole density
constitutes a much smaller correction to the dipole contribution which is almost
negligible outside the reservoirs. This might seem surprising at first given that the results for the quadrupole 
densities agree well for both
summation methods (Fig.~\ref{fig:multicomb}e). The apparent contradiction is 
explained by the fact that the derivatives of the kernels in the evaluation of the potential 
are very different for both methods.
We will get back to this point in Sec.~\ref{subsec:respol} 
when we discuss the macroscopic polarisation.

With regard to the accuracy of the full multipole
approximations (up to the quadrupole term), we observed different trends for the
maximum error of the potential within each expansion. For the slab
expansion we found the maximum error to be about 6 times larger than the error
of the exact potential for the lowest resolution ($N_\text{b}=10$, $\upDelta z = 10.9~\text{\AA}$). 
Upon increasing the resolution, the error ratio approaches unity, which is the
expected behaviour. However, this is not the case for the molecule expansion,
where the error is only about $20\%$ larger than the error of
the exact potential initially, but the difference increases to
about $100\%$ for the highest resolution ($N_\text{b}=3200$, $\upDelta z = 0.034~\text{\AA}$). 
We believe that this behaviour is reasonable, because we never intersect
molecules and cannot resolve the potential inside a molecule correctly. The higher the
 resolution the worse we expect the approximation to become in
the vicinity of the point multipoles. Averaging the potential exactly 
is preferable on all scales, rendering it clearly the method of choice.

\subsection{\label{subsec:resfield}Electrostatic field}
\begin{figure*}
  \centering
   \includegraphics{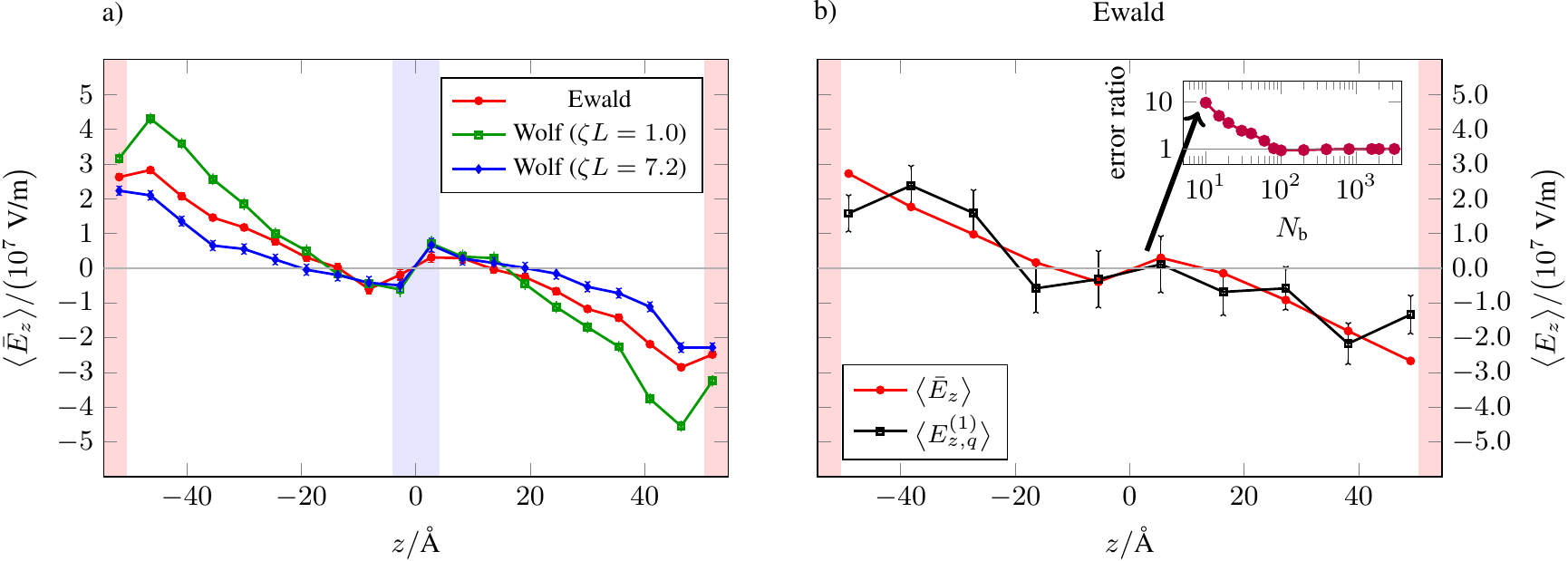}
   \caption{\label{fig:eex}The exact field is shown in figure a) and a
   comparison between the field calculated from the monopole density in the slab
   expansion and the exact result is shown in b). The inset
   compares the ratio of the maximum errors which were calculated from 600 blocks of length $100~\text{ps}$ as a function of
   the number of bins. The arrow indicates the error ratio for the resolution
   shown in the full figure.}
\end{figure*}

The exact results for the field in the sense of Eq.~(\ref{eq:Ez_ex_avg}) are
shown in Fig.~\ref{fig:eex}a.
Focusing on the left half of the
simulation box, we notice that the field is positive and strongest in the vicinity of the hot reservoir. For the peak field
strength we measured values of about
$2.8\times10^7~\text{V/m}$, $4.4\times10^7~\text{V/m}$ and
$2.2\times10^7~\text{V/m}$ for Ewald summation and the Wolf method with weak
and strong damping, respectively. Close to the hot reservoir, the short-ranged method overshoots the Ewald summation result for low damping and vice versa for high damping. We also infer from the figure that the field changes its sign in the vicinity of
the cold reservoir. From the discussion of the potentials in the previous
section (Fig.~\ref{fig:phimulticomb}c) we know that the inversion happens
exactly when the dipole contribution to the field dominates over
the quadrupole contribution.

Comparing our results to the ones reported by Bresme and
co-workers, we find a major discrepancy: In
the original work~\cite{Armstrong2013} the reported fields are about one order 
of magnitude higher than what we found.
Recently, however, it was suggested that the thermally induced
field in a spherical droplet of SPC/E water is of the order of $10^7~\text{V/m}$
after comparison with Ewald summation (PPPM)~\cite{Armstrong2015}. Nevertheless, the
discrepancy still persists as the authors~\cite{Armstrong2015}
suggest that the Wolf method itself is responsible for the overestimated field,
whereas, in fact, the opposite is true for the set of parameters employed in
Ref.~\onlinecite{Armstrong2013}. The Wolf method slightly underestimates the
field if it is calculated consistently, namely using the correct kernel~(see
Fig.~\ref{fig:eex}a). We can reproduce the
results of Armstrong and Bresme closely if we
calculate the field as~\cite{Armstrong2013}
\begin{equation}
\label{eq:EIso}
E_z(z) = 4 \pi \int\limits_{-L_z/2}^z \mathrm dz' \rho_q(z'),
\end{equation}
considering Gaussian units and taking the lower integration bound to be $-L_z/2$
rather than $-\infty$. (A comparison is omitted for brevity.) For Ewald
summation this expression is correct and equivalent to Eq.~(\ref{eq:Ez3Dto1D})
with $G^{'}_\text{1D,E}$ as long as the net dipole density of the box,
\begin{equation}
\label{eq:rhomunet}
  \bar{\rho}_{\mu,L_z}  = \frac{1}{L_z} \int\limits_{-L_z/2}^{L_z/2}
 \mathrm dz' z' \rho_q(z'),
\end{equation}
vanishes. Considering sufficiently long simulations, this is necessarily the
case for our system because of the symmetric setup (see
Figs~\ref{fig:box},~\ref{fig:multicomb}b and d).
If this was not the case, an additional term $ 4 \pi \bar{\rho}_{\mu,L_z}$
would have to be added to the right-hand side of
Eq.~(\ref{eq:EIso}).
The equivalence is trivially shown by rewriting the integral in
Eq.~(\ref{eq:Ez3Dto1D}) taking into account periodicity and charge neutrality.
Alternatively, one can integrate Poisson's equation directly and impose
periodicity by choosing the integration constants accordingly~\cite{Yeh2011}.
However, applying Eq.~(\ref{eq:EIso}) for the Wolf
method is wrong and the discrepancy between our result and the
one of Armstrong and Bresme~\cite{Armstrong2013} can therefore be traced back to
using the incorrect expression in the calculation.

Similarly to what we observed for the potential, considering
exact averages rather than estimating the field from the average charge density yields a huge improvement for low
resolutions. The comparison in
Fig.~\ref{fig:eex}b is carried out for a resolution of $N_\text{b}=10$
($\upDelta z = 10.9~\text{\AA}$) and, as shown in the inset, the error of the
approximative field, i.e. using the negative derivative of $G_\text{1D,E}$ in
Eq.~(\ref{eq:avgphimulticontr}a), is about 10 times larger than the exact
one. For resolutions higher than $N_\text{b}=80~\text{bins}$ ($\upDelta z \leq
1.36~\text{\AA})$, both approaches yield similar errors. Comparing the insets of
Figs~\ref{fig:phiex}b and \ref{fig:eex}b, we notice that the enhancement of the exact method over the approximative one is much higher for
the potential. This can be partly explained by looking at the functional form of
$G^{'}_\text{1D,E}$ (Eq.~(\ref{eqapp:GEders}a) in
Appendix~\ref{secapp:electrostatics}).
The function is piecewise linear and the midpoint rule, which corresponds to multiplying the function value at
the centre of the bin by $\upDelta z$, is exact in the absence of any
discontinuity. Therefore, the advantage of using 
${\bar G}^{'}_\text{1D,E}$ over ${G}^{'}_\text{1D,E}$ for the evaluation of the
field is less significant than for the potential.

\begin{figure*}
  \centering
   \includegraphics{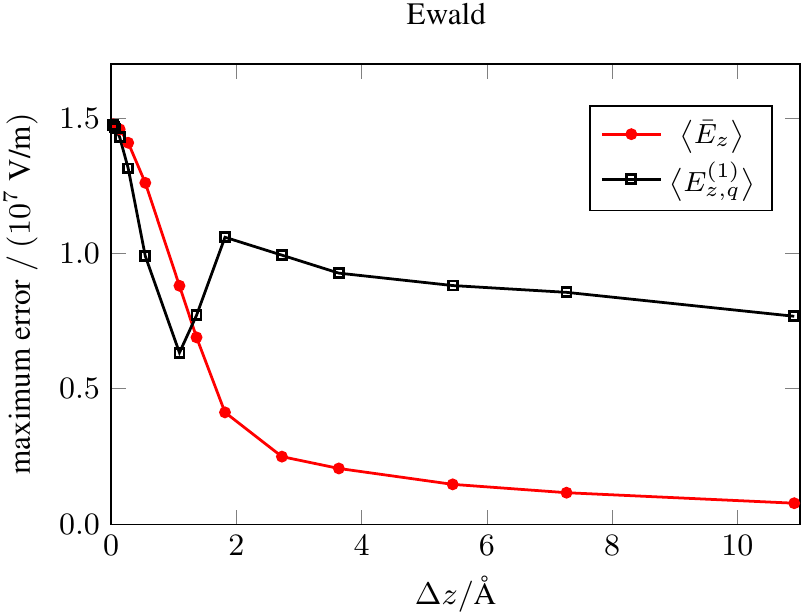}
   \caption{\label{fig:e_error}Spatial maximum
    error as a function of the resolution $\upDelta z$ for the exact field
    (red circles) and the monopole field (black, open squares), respectively.
    (The error is defined as one standard deviation of the mean and the error
    bar in Fig.~\ref{fig:eex}b corresponds to twice the value for $\upDelta z =
    10.9~\text{\AA}$.)}
\end{figure*}

Figure~\ref{fig:e_error} compares the spatial maximum errors for
varying resolutions. Interestingly, for sufficiently high
resolutions of $\upDelta z \leq 1~\text{\AA}$ we found the maximum error of the
approximative method to be up to almost 30\% lower than the one for the exact average. We attribute this to 
 cancellation of errors, since convergence tests support a correct
 implementation. Far more important is the magnitude of the error for high
 resolutions. For simulation time scales of $100~\text{ns}$ the
 error is comparable to the signal itself requiring even longer runs for the
 statistics to be satisfactory.
  Suppose we wanted to get a rough idea of what the field looked like. With the conventional method, i.e. averaging
the charge density first and then integrating it, the best we can do is to calculate
the results on a sufficiently high resolution and then perform some sort of
averaging. On the one hand, this approach is problematic
because the coarse-grained values do not represent the correct bin averages. On
the other hand, it is not straightforward to propagate the statistical errors 
from the fine resolution to the coarse one since the values are
highly correlated. Our proposed method of averaging the potential and the
field analytically eliminates both issues and yields a huge improvement for low
resolutions reducing the required simulation time scales by up to two orders of 
magnitude for the same quality of statistics.

\subsection{\label{subsec:respol}Macroscopic polarisation}
Our final goal in this section is to relate the molecular
multipole densities to the macroscopic polarisation. We show that the macroscopic
Maxwell equation
\begin{equation}
\label{eq:MaxwellEDP}
{\bar E}_z(z) = {\bar D}_z(z) - 4 \pi {\bar P}_z(z)
\end{equation}
holds locally for the bin averages calculated with Ewald summation, where ${\bar
P}_z$ and ${\bar D}_z$ are the $z$-components of polarisation and displacement field, respectively.
We do not make any a priori assumptions about the locality~\cite{Neumann1986a} 
and use the multipole expansions developed in Sec.~\ref{sec:multi} as a general
starting point for the discussion. We then identify the quantities on the
right-hand side of Eq.~(\ref{eq:MaxwellEDP}) after simplifying the
expressions. We note that our analysis only holds in the context of sufficiently long simulations (like in Sec.~\ref{sec:multi}), because
we use $\rho_q(z)$ in place of the full $\rho_q(\mathbold r)$.
This simplifies the discussion in that we only need to consider the $z$-component of the spatially
averaged dipole density, $\bar{\rho}_\mu$, and
the density of $Q_{zz}$ given by $\bar{\rho}_Q$, respectively.

The water molecules comprise the polarisable background medium and
there are no free charges. From our discusion in Sec.~\ref{subsec:respot}, we
know that the dipole contribution alone yields a poor approximation to the
potential (Figs~\ref{fig:phimulticomb}c and d). 
As a natural extension we
considered the quadrupole contribution~\cite{Boettcher1973}, which
was also found to be important in simulation studies of interfacial electric
fields~\cite{Wilson1989, Glosli1996, Sokhan1997}.
With the inclusion of this contribution, the potentials from the molecular
multipole expansions match the exact potentials very well for both methods,
respectively. The corresponding expression for the field extends to 
\begin{subequations}
\label{eq:EfromPquad}
\begin{alignat}{2}
{\bar E}_z(z) &= && \int\limits_{-\frac{L_z}{2}}^{\frac{L_z}{2}} \mathrm dz'\ \bigg[ G_\text{1D}^{''}(z-z') {\bar \rho}_{\mu}(z') 
                   -G_\text{1D}^{'''}(z-z')  {\bar \rho}_{Q}(z')\bigg] \\
              &= &&\int\limits_{-\frac{L_z}{2}}^{\frac{L_z}{2}} \mathrm dz'\ G_\text{1D}^{''}(z-z')
                                     \left[ {\bar \rho}_{\mu}(z') - {\bar \rho}^{'}_{Q}(z')  \right] ,
\end{alignat}
\end{subequations}
where the derivatives of the kernels are given in
Appendix~\ref{secapp:electrostatics}. To get to Eq.~(\ref{eq:EfromPquad}b) we
integrated the second term in Eq.~(\ref{eq:EfromPquad}a) by parts taking into account the periodicity. 
We can solve the above integral analytically for Ewald summation and find that
\begin{equation}
 \label{eq:EPD}
 {\bar E}_z(z) = -4 \pi \left[{\bar\rho}_{\mu}(z) - {\bar\rho}_{\mu,L_z} -
 {\bar \rho}^{'}_Q(z) \right],
 \end{equation}
where ${\bar \rho}_{\mu, L_z}$ is the box average of ${\bar \rho}_{\mu}(z)$.
In general, we can identify this contribution with ${\bar D}_z$ as it
corresponds to the (constant) field arising from an induced surface 
charge density at infinity (tin-foil boundary
conditions). We refer to Refs~\onlinecite{Stengel2008} and
\onlinecite{Zhang2016} for a more general discussion. Although the
instantaneous value of ${\bar D}_z$ may fluctuate, we know that its time average vanishes, 
because our system does not exhibit a net dipole moment (Figs~\ref{fig:multicomb}b and d).
For Ewald summation the definition of polarisation as
\begin{equation}
  \label{eq:PQuad}
  {\bar P}_z(z) = {\bar \rho}_{\mu}(z)-{\bar \rho}^{'}_Q(z)
\end{equation}
therefore naturally leads to the correct proportionality of
$\langle {\bar P}_z(z)\rangle = -\langle {\bar
E}_z(z)\rangle/4\pi$. For the Wolf method the relation between electric field
and polarisation (as defined above) is more complicated, because we cannot solve the integral in Eq.~(\ref{eq:EfromPquad}b) analytically. More
importantly, we cannot expect the short-ranged method to predict fields
accurately in general, because its kernel is not a solution of Poisson's equation. 
The estimates for the thermally induced fields might be reasonable, but it is
trivial to come up with an example, such as a plate capacitor, for which the
method would fail.

Finally, we would like to discuss the macroscopic Maxwell equation
(\ref{eq:MaxwellEDP}) in the context of the slab expansion. As shown in
Figs~\ref{fig:phimulticomb}a-b, we can identify all relevant multipole
contributions to the potential and recover a good approximation to the
exact solution implying overall consistency. Due to the nature of the spatial
averaging, we obtain a non-vanishing charge density
(Fig.~\ref{fig:multicomb}a) for our inhomogeneous system. 
This is inconsistent, however, with the derivation of Eq.~(\ref{eq:MaxwellEDP}),
where charges within a molecule are summed first in order to get from the
microscopic to the macroscopic description~\cite{Jackson1998, Boettcher1973}
and the charge density vanishes identically. Identification of displacement field
and polarisation is therefore not meaningful for the slab expansion. This
problem is avoided altogether in the molecule expansion, which is consistent with Eq.~(\ref{eq:MaxwellEDP}), and we can unambiguously identify all terms in the macroscopic Maxwell equation.

\section{\label{sec:concl}Conclusions}
In this paper we have analysed the electric fields and
multipole moments induced by a strong thermal gradient in NEMD simulations
of water in a setup which was previously studied by Armstrong
and Bresme~\cite{Armstrong2013}.
Our comparison comprises results for two different treatments of Coulomb
interactions, namely Ewald summation and the short-ranged Wolf method. 
The latter was employed in most of the previous studies
on the thermo-polarisation effect~\cite{Bresme2008,Muscatello2011a,
Armstrong2013, Armstrong2013a, Armstrong2015}.
We identified two key differences to the
literature data: Firstly, the Wolf method fails to reproduce the dipole density
correctly for parameters that work well in equilibrium simulations. 
The molecules point, on average, in opposite directions as
compared to Ewald summation and the alignment is strongly
enhanced.

Secondly, for both methods the peak field
strength is of the order of $10^7~\text{V/m}$. However, for the Wolf method the
result depends sensitively on the employed parameters. For low damping the Wolf
method slightly overestimates the field obtained with Ewald summation 
and vice versa for high damping. The results are therefore in direct constrast
to very recent findings of Bresme and co-workers~\cite{Armstrong2015} who
reported that the Wolf method overestimates the field by an order of
magnitude. In fact, we argue that the employed formula for the calculation of
the field is incorrect. Taking such truncation into account correctly results in
comparable results for the electric field.

Another key result of this paper are the highly improved spatial averages of the
potential and the field for low resolutions. We propose
to integrate these quantities analytically over the bins rather than calculating
them from the time-averaged charge density, as is usually done in the
literature. Potentials and fields then truly represent the exact spatial
averages over the microscopic or macroscopic control volumes.
We showed that this procedure is straightforward for both summation methods and requires no computational overhead. 
Comparing the ratio of maximum errors, we found a more than
20-fold reduction of the error for the potential and a 10-fold
reduction for the field at the coarsest resolution of $\upDelta z \approx
10.9~\text{\AA}$. Consequently, employing the new method can reduce the simulation time scales by up to two orders
of magnitude for the same quality of statistics. The advantage of calculating analytical 
averages becomes less significant with
increasing spatial resolution and both methods are comparable for
resolutions of $\upDelta z \leq 1~\text{\AA}$. However, in this
case the magnitude of the statistical error is comparable to the signal itself rendering
the results meaningless.

In addition, we found that accurate estimates of the potential and the
field can be obtained by approximating the water molecules as ideal point
dipoles and quadrupoles. For low spatial resolutions we found this approach to yield
considerably better results than the calculation from the averaged charge
density. Our detailed comparison of the results for the slab and molecule
expansions illustrates that the ratio of the individual contributions
depends on the control volume we choose for the expansion. For slabs almost all
the information can be recovered by considering the monopole, as is usually done
in the literature.
However, in the molecule expansion the dipole and the quadrupole contributions
are significant and both have to be considered in order to recover results
from the exact calculation accurately.

Finally, taking into account the quadrupole contribution leads to the expected
proportionality between the polarisation and the macroscopic Maxwell field in
accordance with the macroscopic Maxwell equations. The Wolf method fails to
satisfy this relation entirely. Based on its shortcomings, we therefore conclude that the
method is not suitable for reproducing the electrostatic key quantities 
in inhomogeneous systems reliably. This is in agreement with
the findings of Takahashi and co-workers~\cite{Takahashi2011}, who
reported poor predictions for the electrostatic potential and dipolar
orientations in simulations of the liquid--vapour interface, even for
cutoff radii almost six times larger than the maximum value considered in
this work.
 
\begin{acknowledgments}
The authors should like to dedicate this paper
to the memory of Simon de Leeuw, who was a pioneer in the calculation of Coulomb
effects in simulations. P.~W. would like to thank the Austrian Academy of
Sciences for financial support through a DOC Fellowship, and for covering the travel expenses for the CECAM
workshop in Zaragoza in May 2015, where these results were first presented.
P.~W. would also like to thank Chao Zhang
for pointing out the equivalence of the two expressions
for the electric field discussed in Sec.~\ref{subsec:resfield}, 
Michiel Sprik for emphasising the importance of the
quadrupole contribution in experimental studies of interfacial systems, as well
as Aleks Reinhardt and other members of the Frenkel and Dellago groups for their
advice. We further acknowledge support from the Federation of Austrian
Industry (IV) Carinthia (P.~W.), the University of Zagreb and Erasmus
SMP (D. Fijan), the Human Frontier Science Program and
Emmanuel College (A.~\v{S}.), the Austrian Science
Fund FWF within the SFB Vicom project F41 (C.~D.), and the Engineering and
Physical Sciences Research Council Programme Grant EP/I001352/1 (D.~F.).
Additional data related to this publication are available at the University of
Cambridge data repository (\href{http://dx.doi.org/10.17863/CAM.118}{http://dx.doi.org/10.17863/CAM.118}).
\end{acknowledgments}
\appendix

\section{\label{secapp:electrostatics}Electrostatics}
\subsection{\label{subsecapp:wolf}Wolf method}
Our goal is to integrate $G_\text{W}$ over the entire cutoff sphere in order to
obtain ${G}_\text{1D,W}$. To this end we have to evaluate the integral 
\begin{subequations}
\begin{align}
{G}_\text{1D,W}(z)     &= \int\limits_{-\frac{L_x}{2}}^{\frac{L_x}{2}} \mathrm dx \
\int\limits_{-\frac{L_y}{2}}^{\frac{L_y}{2}} \mathrm dy \ G_{\text{W}}(x, y, z) \label{eq:GW1Da} \\
                         &= 2\pi\int\limits_{0}^{s_\text{c}(z)} \mathrm ds \ s
                         \left[ \frac{\text{erfc}(\zeta \sqrt{s^2+z^2} )}{\sqrt{s^2+z^2}}  -  \frac{\text{erfc}(\zeta r_\text{c})}{r_\text{c}}  \right] \label{eq:GW1Db},
\end{align}
\end{subequations}
where $r^2 = x^2+y^2+z^2 = s^2 + z^2$. We first consider the integral
\begin{align}
I(z) = \int\limits_0^{s_\text{c}(z)} \mathrm ds\ s \frac{\erfc(\zeta \sqrt{s^2 + z^2})}{\sqrt{s^2+z^2}} \label{eqA:GWint}
\end{align}
and use the substitution $\tau(s,z) = \sqrt{s^2+z^2}$ to rewrite the expression as
\begin{align}
I(z) = \int\limits_{\tau(0,z)}^{\tau(s_\text{c}(z),z)} \mathrm d\tau\ \erfc(\zeta \tau).
\end{align}
Using integration by parts it is easy to show that the result is
\begin{align}
I(z) = r_\text{c} \erfc(\zeta r_\text{c}) - |z|\erfc(\zeta|z|) 
        + \frac{ \text{e}^{-\zeta^2 z^2} - \text{e}^{-\zeta^2 r_\text{c}^2} }{\sqrt \pi \zeta}
\end{align}
for $|z|\leq r_\text{c}$ and zero otherwise. The integration of the second 
term in Eq.~(\ref{eq:GW1Db}) is trivial and the averaged kernel is
given by
\begin{alignat}{2}
\frac{{G}_\text{1D,W}(z) }{2\pi}
     &=&& \phantom{{}+{}}  \frac{r_\text{c}}{2} \erfc(\zeta r_\text{c}) - |z| \erfc(\zeta |z|) \label{eq:GWavgn}  \\
     & && + \frac{  \text{e}^{-\zeta^2 z^2} - \text{e}^{-\zeta^2 r_\text{c}^2}} {\sqrt \pi \zeta}
             + \frac{z^2 \erfc(\zeta r_\text{c})}{2 r_\text{c}} \nonumber 
\end{alignat}
for $|z|\leq r_\text{c}$ and it vanishes otherwise. The first three
derivatives of this function are  
\begin{subequations}
\label{eqapp:GWders}
\begin{alignat}{2}
\frac{{G}^{'}_\text{1D,W}(z)}{2\pi} &=&& - \sgn(z) \erfc(\zeta |z|) +
		\frac{z \erfc(\zeta r_\text{c})} {r_\text{c}},
\\
\frac{{G}^{''}_\text{1D,W}(z)}{2\pi} &=&&  -2 \delta(z) \erfc(\zeta|z|) \\
                        & && + \frac{2\zeta}{\sqrt \pi} \sgn(z)^2 \text{e}^{-\zeta^2 z^2} + \frac{\erfc(\zeta
				r_\text{c})}{r_\text{c}}  \nonumber ,
\\
\frac{{G}^{'''}_\text{1D,W}(z)}{2\pi} &=&&   -2 \delta^{'}(z)
\erfc(\zeta|z|)  \\
                   & && + \frac{2\zeta}{\sqrt \pi} \sgn(z) \text{e}^{-\zeta^2 z^2}
\Big[ -2 \zeta^2 |z| + 6 \delta(z) \Big]   \nonumber ,
\end{alignat}
\end{subequations}
respectively.

\subsection{\label{subsecapp:ewald}Ewald summation}
Instead of integrating the kernel $G_\text{E}$ (Eq.~(\ref{eq:GEsim}))
directly, we replace it by $G_\text{PBC}$ (Eq.~(\ref{eq:Gpbc3D})) in order
to simplify the problem. The sum in Eq.~(\ref{eq:Gpbc3D}) is only conditionally
convergent, which is why we are formally not allowed to change the
order of integration and summation. However, if we ignore this fact we arrive at
the same result that we would have obtained by considering $G_\text{E}$
directly. This yields
\begin{subequations}
\label{eqapp:GE1D}
\begin{align}
{G}_\text{1D,E}(z) &= \int\limits_{-\frac{L_x}{2}}^{\frac{L_x}{2}}  \mathrm dx \
\int\limits_{-\frac{L_y}{2}}^{\frac{L_y}{2}}  \mathrm dy \ G_\text{PBC}(x, y, z)  \\
                          &= \int\limits_{-\frac{L_x}{2}}^{\frac{L_x}{2}}  \mathrm dx \ \int\limits_{-\frac{L_y}{2}}^{\frac{L_y}{2}}  \mathrm dy  
                             \frac{1}{L_x L_y L_z} \sum_{\mathbold k \neq \mathbf 0}
                             \frac{4\pi}{k^2} \text{e}^{i \mathbold k \cdot \mathbold r}  \\  
                           &= \frac{1}{L_z} \sum_{k_z \neq  0}
                             \frac{4\pi}{k_z^2} \text{e}^{i k_z z}.  
\end{align}
\end{subequations}
In the last step, we make use of the fact that the integration eliminates all terms in the summation 
for which $k_x \neq 0$ or $k_y \neq 0$.
The inverse Fourier transform in Eq.~(\ref{eqapp:GE1D}c) is
\begin{align}
{G}_\text{1D,E}(z)= 2 \pi \left(-|z| + \frac{{z}^2}{L_z} + \frac{L_z}{6} \right) \label{eqapp:GE1Dreal}
\end{align}
and the first three derivatives of this expression are given by
\begin{subequations}
\label{eqapp:GEders}
\begin{align}
{G}^{'}_\text{1D,E}(z) &=  2\pi \left(- \sgn(z)  + \frac{2 z}{L_z} \right),\\
{G}^{''}_\text{1D,E}(z) &=  2\pi \left( -2 \delta(z) + \frac{2}{L_z}
\right), \\
{G}^{'''}_\text{1D,E}(z) &= -4 \pi  \delta^{'}(z),
\end{align}
\end{subequations}
respectively.

\subsection{\label{subsecapp:exactavg}Exact averaging}
\begin{figure}[]
  \centering
   \includegraphics{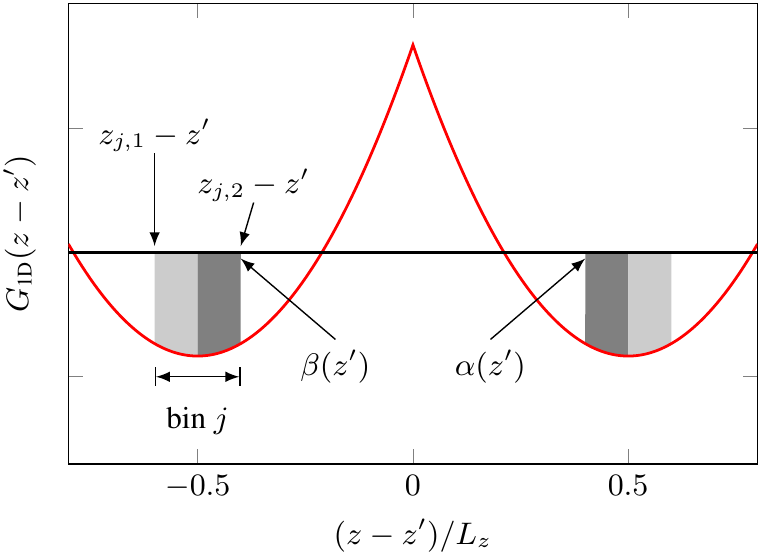}
   \caption{Integration of the spatially averaged kernel for the
   case where the separation of the charge at $z'$ and the bin $j$ covering the interval
   $[z_{j,1}, z_{j,2}]$ is such that periodicity has to be taken into account.
   $\alpha_j(z')$ and $\beta_j(z')$ are the nearest images of the bin
   boundaries.}
  \label{fig:g1de}
\end{figure}

The aim is to average
the one-dimensional kernel analytically for any bin $j$ 
of width $\upDelta z = z_{j,2} - z_{j,1}$ to
obtain
\begin{align}
\label{eqapp:G1DAnaAvg}
{\bar G}_{\text{1D},j}(z') = & \frac{1}{\upDelta z} 
\int\limits_{z_{j,1}}^{z_{j,2}} \mathrm dz\ {G_\text{1D}}(z-z')
\end{align}
taking into account the periodicity. As mentioned in
Sec.~\ref{sec:electrostatics}, in our notation we understand the argument $z-z'$ to be mapped 
back to the interval $[-\frac{L_z}{2},\frac{L_z}{2}]$ implicitly.
The interesting case, where the separation of the charge at $z'$ and the
bin covering the interval $[z_{j,1}, z_{j,2}]$ is such that periodicity
has to be taken into account in the integration, is illustrated in
Fig.~\ref{fig:g1de}.

The first step is to map the distances from $z'$ to the bin boundaries back
into the reference interval using the function
\begin{equation}
\pbc(z) = z - L_z \round\left(\frac{z}{L_z} \right),
\end{equation}
where $\round(z)$ gives the nearest integral number to $z$. Applying this 
function yields the shortest distances to the nearest images
which we label with
\begin{subequations}
\label{eqapp:pbczjl}
\begin{align}
\alpha_j(z')=\pbc(z_{j,1}-z'), \\
\beta_j(z')=\pbc(z_{j,2}-z'), 
\end{align}
\end{subequations}
respectively. For the case shown in Fig.~\ref{fig:g1de}, where $\beta_j(z') <
\alpha_j(z')$, we can split the original expression into the two integrals
\begin{equation}
\label{eqapp:G1DAnaAvgSplit}
{\bar G}_{\text{1D},j}(z') = \frac{1}{\upDelta z}
\left[\int\limits_{-\frac{L_z}{2}}^{\beta_j(z')} \mathrm dz\
{G_\text{1D}}(z) + \int\limits_{\alpha_j(z')}^{\frac{L_z}{2}} \mathrm dz\
{G_\text{1D}}(z) \right].
\end{equation}
In order to simplify the integration further, we focus on the case of Ewald
summation. Application of the procedure to the Wolf method is omitted for
brevity, because the integration is tedious. We know that the average of
$G_\text{1D,E}$ over the reference interval vanishes because 
the term corresponding to $k_z=0$ in Eq.~(\ref{eqapp:GE1D}c) is absent. Therefore, the special
case shown in Fig.~\ref{fig:g1de} reduces to the ordinary case 
\begin{equation}
\label{eqapp:G1DAnaAvgRecomb}
{\bar G}_{\text{1D,E},j}(z') = \frac{1}{\upDelta z}
\int\limits_{\alpha_j(z')}^{\beta_j(z')} \mathrm dz\
{G_\text{1D,E}}(z),
\end{equation}
in which the entire bin is located inside the reference box. All possible
scenarios are therefore taken into account by straightforward integration of
Eq.~(\ref{eqapp:GE1Dreal}), which yields
\begin{alignat}{2}
\label{eqapp:G1DAnaAvgRes}
\frac{ {\bar G}_{\text{1D,E},j}(z')}{2\pi} &=  
 && \phantom{{}+{}} \frac{  \alpha_j(z') |\alpha_j(z')| -\beta_j(z') |\beta_j(z')|  
 }{2\upDelta z}  \nonumber \\
 & &&+ \frac{\beta^3_j(z')-\alpha^3_j(z')}{3 L_z \upDelta z}  \\
 & && + \frac{L_z \big(\beta_j(z')-\alpha_j(z')\big)}{6 \upDelta z} \nonumber.
\end{alignat}
Likewise, we find
\begin{equation}
\label{eqapp:G1DerAnaAvgRes}
\frac{{\bar G}^{'}_{\text{1D,E},j}(z')}{2\pi} =   
  \frac{|\alpha_j{(z')}| - |\beta_j(z')|}{\upDelta z} + \frac{\beta^2_j(z') - \alpha^2_j(z')}{\upDelta z L_z}
\end{equation}
for the average of the derivative. Equations~(\ref{eqapp:G1DAnaAvgRes}) and
(\ref{eqapp:G1DerAnaAvgRes}) along with Eqs~(\ref{eqapp:pbczjl}a--b) can 
be substituted into Eqs~(\ref{eq:phi_ex_avg}) and (\ref{eq:Ez_ex_avg}) to
calculate the exact averages of the potential and
 the field, respectively.

\section{\label{secapp:valid}Validation}
In this section, we compare the oxygen-oxygen pair correlation function, $g(r)$, 
the oxygen-oxygen velocity autocorrelation function, $\text{VACF}(t)$, 
a cumulative estimate of the dielectric constant, $\epsilon(t)$, 
and the distance-dependent Kirkwood $g$-factor, $G_\text{K}(r)$.
We refer to Refs~\onlinecite{Frenkel2002} and \onlinecite{Neumann1986} for a detailed
discussion and the relevant formulae. All quantities
were sampled during 2~\text{ns} \textit{NVE} simulations before
imposing the temperature gradients.
\begin{figure*}
 \centering
   \includegraphics{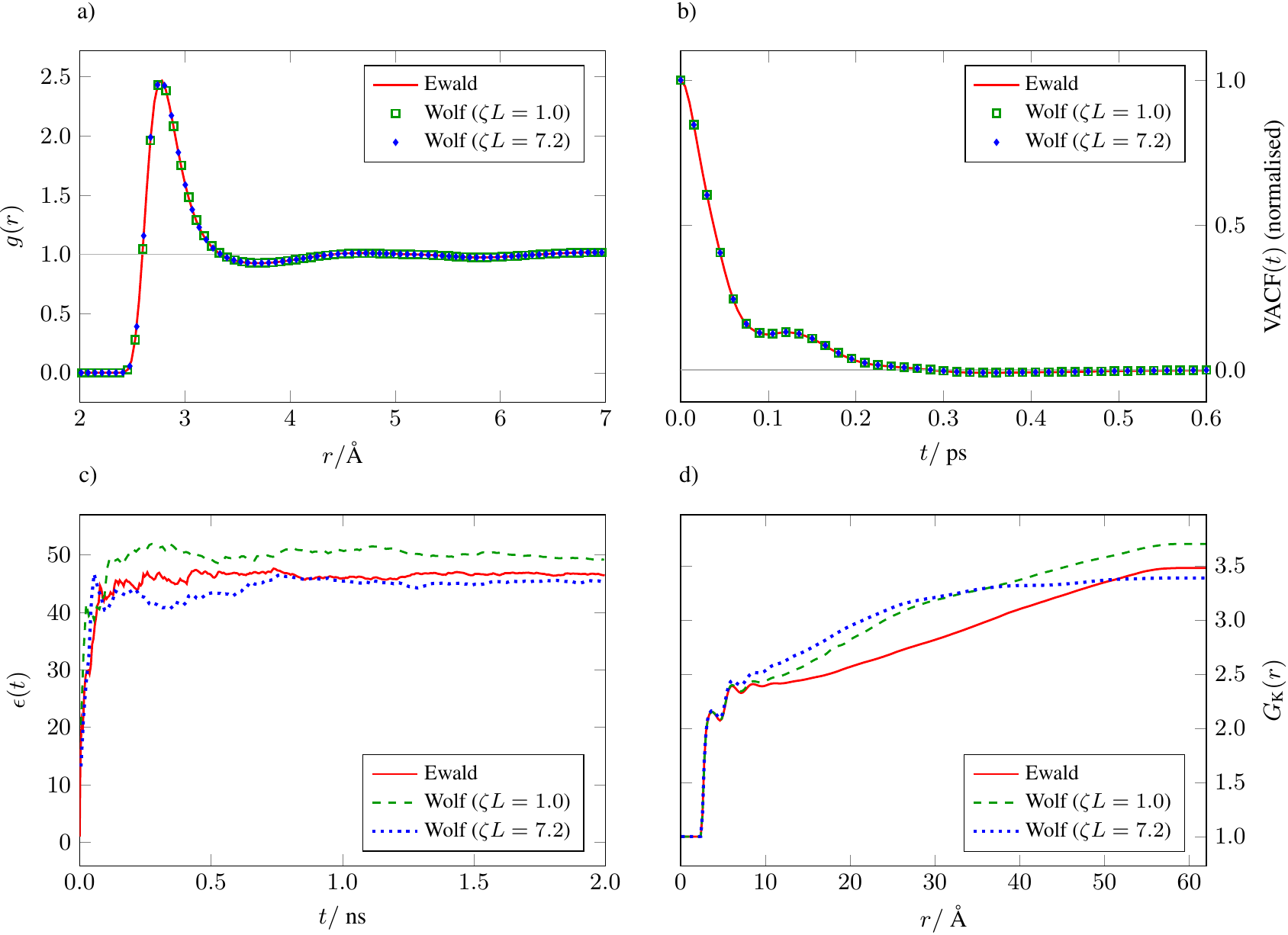}
    \caption{\label{fig:validcomb}The comparison comprises (a) the
    oxygen-oxygen pair correlation function, (b) the oxygen-oxygen VACF, (c) a
    cumulative estimate of the dielectric constant and (d) the
    distance-dependent Kirkwood $g$-factor.}
\end{figure*}
The results are shown in Figs~\ref{fig:validcomb}a--d. As we can see, all sets
of parameters lead to excellent agreement for $g(r)$ and $\text{VACF}(t)$
(Figs~\ref{fig:validcomb}a--b). The dielectric constant
(Fig.~\ref{fig:validcomb}c) is well reproduced by the Wolf method with
strong damping, whereas weak damping leads to an overestimation. More insight
about the structural properties can be gained by looking at $G_\text{K}(r)$
in Fig.~\ref{fig:validcomb}d. For very short distances both sets of
parameters for the Wolf method yield a good agreement with Ewald summation. We
note that for the weak damping the agreement extends a bit further than for
strong damping, which is consistent with our
observations for the model system. We also note that the shape of $G_\text{K}(r)$ looks different
for our elongated box as compared to a cubic box.

\nocite{*}


\begin{thebibliography}{60}%
\makeatletter
\providecommand \@ifxundefined [1]{%
 \@ifx{#1\undefined}
}%
\providecommand \@ifnum [1]{%
 \ifnum #1\expandafter \@firstoftwo
 \else \expandafter \@secondoftwo
 \fi
}%
\providecommand \@ifx [1]{%
 \ifx #1\expandafter \@firstoftwo
 \else \expandafter \@secondoftwo
 \fi
}%
\providecommand \natexlab [1]{#1}%
\providecommand \enquote  [1]{``#1''}%
\providecommand \bibnamefont  [1]{#1}%
\providecommand \bibfnamefont [1]{#1}%
\providecommand \citenamefont [1]{#1}%
\providecommand \href@noop [0]{\@secondoftwo}%
\providecommand \href [0]{\begingroup \@sanitize@url \@href}%
\providecommand \@href[1]{\@@startlink{#1}\@@href}%
\providecommand \@@href[1]{\endgroup#1\@@endlink}%
\providecommand \@sanitize@url [0]{\catcode `\\12\catcode `\$12\catcode
  `\&12\catcode `\#12\catcode `\^12\catcode `\_12\catcode `\%12\relax}%
\providecommand \@@startlink[1]{}%
\providecommand \@@endlink[0]{}%
\providecommand \url  [0]{\begingroup\@sanitize@url \@url }%
\providecommand \@url [1]{\endgroup\@href {#1}{\urlprefix }}%
\providecommand \urlprefix  [0]{URL }%
\providecommand \Eprint [0]{\href }%
\providecommand \doibase [0]{http://dx.doi.org/}%
\providecommand \selectlanguage [0]{\@gobble}%
\providecommand \bibinfo  [0]{\@secondoftwo}%
\providecommand \bibfield  [0]{\@secondoftwo}%
\providecommand \translation [1]{[#1]}%
\providecommand \BibitemOpen [0]{}%
\providecommand \bibitemStop [0]{}%
\providecommand \bibitemNoStop [0]{.\EOS\space}%
\providecommand \EOS [0]{\spacefactor3000\relax}%
\providecommand \BibitemShut  [1]{\csname bibitem#1\endcsname}%
\let\auto@bib@innerbib\@empty
%</preamble>
\bibitem [{\citenamefont {Doktycz}\ and\ \citenamefont
  {Suslick}(1990)}]{Doktycz1990}%
  \BibitemOpen
  \bibfield  {author} {\bibinfo {author} {\bibfnamefont {S.~J.}\ \bibnamefont
  {Doktycz}}\ and\ \bibinfo {author} {\bibfnamefont {K.~S.}\ \bibnamefont
  {Suslick}},\ }\href {\doibase 10.1126/science.2309118} {\bibfield  {journal}
  {\bibinfo  {journal} {Science}\ }\textbf {\bibinfo {volume} {247}},\ \bibinfo
  {pages} {1067} (\bibinfo {year} {1990})}\BibitemShut {NoStop}%
\bibitem [{\citenamefont {Govorov}\ and\ \citenamefont
  {Richardson}(2007)}]{Govorov2007}%
  \BibitemOpen
  \bibfield  {author} {\bibinfo {author} {\bibfnamefont {A.~O.}\ \bibnamefont
  {Govorov}}\ and\ \bibinfo {author} {\bibfnamefont {H.~H.}\ \bibnamefont
  {Richardson}},\ }\href {\doibase 10.1016/S1748-0132(07)70017-8} {\bibfield
  {journal} {\bibinfo  {journal} {Nano Today}\ }\textbf {\bibinfo {volume}
  {2}},\ \bibinfo {pages} {30} (\bibinfo {year} {2007})}\BibitemShut {NoStop}%
\bibitem [{\citenamefont {Bresme}\ \emph {et~al.}(2008)\citenamefont {Bresme},
  \citenamefont {Lervik}, \citenamefont {Bedeaux},\ and\ \citenamefont
  {Kjelstrup}}]{Bresme2008}%
  \BibitemOpen
  \bibfield  {author} {\bibinfo {author} {\bibfnamefont {F.}~\bibnamefont
  {Bresme}}, \bibinfo {author} {\bibfnamefont {A.}~\bibnamefont {Lervik}},
  \bibinfo {author} {\bibfnamefont {D.}~\bibnamefont {Bedeaux}}, \ and\
  \bibinfo {author} {\bibfnamefont {S.}~\bibnamefont {Kjelstrup}},\ }\href
  {\doibase 10.1103/PhysRevLett.101.020602} {\bibfield  {journal} {\bibinfo
  {journal} {Phys. Rev. Lett.}\ }\textbf {\bibinfo {volume} {101}},\ \bibinfo
  {pages} {020602} (\bibinfo {year} {2008})}\BibitemShut {NoStop}%
\bibitem [{\citenamefont {Muscatello}\ \emph {et~al.}(2011)\citenamefont
  {Muscatello}, \citenamefont {R\"{o}mer}, \citenamefont {Sala},\ and\
  \citenamefont {Bresme}}]{Muscatello2011a}%
  \BibitemOpen
  \bibfield  {author} {\bibinfo {author} {\bibfnamefont {J.}~\bibnamefont
  {Muscatello}}, \bibinfo {author} {\bibfnamefont {F.}~\bibnamefont
  {R\"{o}mer}}, \bibinfo {author} {\bibfnamefont {J.}~\bibnamefont {Sala}}, \
  and\ \bibinfo {author} {\bibfnamefont {F.}~\bibnamefont {Bresme}},\ }\href
  {\doibase 10.1039/c1cp21895f} {\bibfield  {journal} {\bibinfo  {journal}
  {Phys. Chem. Chem. Phys.}\ }\textbf {\bibinfo {volume} {13}},\ \bibinfo
  {pages} {19970} (\bibinfo {year} {2011})}\BibitemShut {NoStop}%
\bibitem [{\citenamefont {R\"{o}mer}\ \emph {et~al.}(2012)\citenamefont
  {R\"{o}mer}, \citenamefont {Bresme}, \citenamefont {Muscatello},
  \citenamefont {Bedeaux},\ and\ \citenamefont {Rub\'{\i}}}]{Romer2012}%
  \BibitemOpen
  \bibfield  {author} {\bibinfo {author} {\bibfnamefont {F.}~\bibnamefont
  {R\"{o}mer}}, \bibinfo {author} {\bibfnamefont {F.}~\bibnamefont {Bresme}},
  \bibinfo {author} {\bibfnamefont {J.}~\bibnamefont {Muscatello}}, \bibinfo
  {author} {\bibfnamefont {D.}~\bibnamefont {Bedeaux}}, \ and\ \bibinfo
  {author} {\bibfnamefont {J.~M.}\ \bibnamefont {Rub\'{\i}}},\ }\href {\doibase
  10.1103/PhysRevLett.108.105901} {\bibfield  {journal} {\bibinfo  {journal}
  {Phys. Rev. Lett.}\ }\textbf {\bibinfo {volume} {108}},\ \bibinfo {pages}
  {105901} (\bibinfo {year} {2012})}\BibitemShut {NoStop}%
\bibitem [{\citenamefont {Armstrong}\ and\ \citenamefont
  {Bresme}(2013)}]{Armstrong2013}%
  \BibitemOpen
  \bibfield  {author} {\bibinfo {author} {\bibfnamefont {J.~A.}\ \bibnamefont
  {Armstrong}}\ and\ \bibinfo {author} {\bibfnamefont {F.}~\bibnamefont
  {Bresme}},\ }\href {\doibase 10.1063/1.4811291} {\bibfield  {journal}
  {\bibinfo  {journal} {J. Chem. Phys.}\ }\textbf {\bibinfo {volume} {139}},\
  \bibinfo {pages} {014504} (\bibinfo {year} {2013})}\BibitemShut {NoStop}%
\bibitem [{\citenamefont {Armstrong}, \citenamefont {Lervik},\ and\
  \citenamefont {Bresme}(2013)}]{Armstrong2013a}%
  \BibitemOpen
  \bibfield  {author} {\bibinfo {author} {\bibfnamefont {J.}~\bibnamefont
  {Armstrong}}, \bibinfo {author} {\bibfnamefont {A.}~\bibnamefont {Lervik}}, \
  and\ \bibinfo {author} {\bibfnamefont {F.}~\bibnamefont {Bresme}},\ }\href
  {\doibase 10.1021/jp408485d} {\bibfield  {journal} {\bibinfo  {journal} {J.
  Phys. Chem. B}\ }\textbf {\bibinfo {volume} {117}},\ \bibinfo {pages} {14817}
  (\bibinfo {year} {2013})}\BibitemShut {NoStop}%
\bibitem [{\citenamefont {De~Groot}\ and\ \citenamefont
  {Mazur}(1962)}]{Groot2013}%
  \BibitemOpen
  \bibfield  {author} {\bibinfo {author} {\bibfnamefont {S.~R.}\ \bibnamefont
  {De~Groot}}\ and\ \bibinfo {author} {\bibfnamefont {P.}~\bibnamefont
  {Mazur}},\ }\href@noop {} {\emph {\bibinfo {title} {{Non-equilibrium
  thermodynamics}}}},\ \bibinfo {edition} {1st}\ ed.\ (\bibinfo  {publisher}
  {North-Holland Publishing Company},\ \bibinfo {address} {Amsterdam},\
  \bibinfo {year} {1962})\BibitemShut {NoStop}%
\bibitem [{\citenamefont {Ewald}(1921)}]{Ewald1921}%
  \BibitemOpen
  \bibfield  {author} {\bibinfo {author} {\bibfnamefont {P.~P.}\ \bibnamefont
  {Ewald}},\ }\href@noop {} {\bibfield  {journal} {\bibinfo  {journal} {Ann.
  Phys.}\ }\textbf {\bibinfo {volume} {64}},\ \bibinfo {pages} {253} (\bibinfo
  {year} {1921})}\BibitemShut {NoStop}%
\bibitem [{\citenamefont {Fennell}\ and\ \citenamefont
  {Gezelter}(2006)}]{Fennell2006}%
  \BibitemOpen
  \bibfield  {author} {\bibinfo {author} {\bibfnamefont {C.~J.}\ \bibnamefont
  {Fennell}}\ and\ \bibinfo {author} {\bibfnamefont {J.~D.}\ \bibnamefont
  {Gezelter}},\ }\href {\doibase 10.1063/1.2206581} {\bibfield  {journal}
  {\bibinfo  {journal} {J. Chem. Phys.}\ }\textbf {\bibinfo {volume} {124}},\
  \bibinfo {pages} {234104} (\bibinfo {year} {2006})}\BibitemShut {NoStop}%
\bibitem [{\citenamefont {Wolf}\ \emph {et~al.}(1999)\citenamefont {Wolf},
  \citenamefont {Keblinski}, \citenamefont {Phillpot},\ and\ \citenamefont
  {Eggebrecht}}]{Wolf1999}%
  \BibitemOpen
  \bibfield  {author} {\bibinfo {author} {\bibfnamefont {D.}~\bibnamefont
  {Wolf}}, \bibinfo {author} {\bibfnamefont {P.}~\bibnamefont {Keblinski}},
  \bibinfo {author} {\bibfnamefont {S.~R.}\ \bibnamefont {Phillpot}}, \ and\
  \bibinfo {author} {\bibfnamefont {J.}~\bibnamefont {Eggebrecht}},\ }\href
  {\doibase 10.1063/1.478738} {\bibfield  {journal} {\bibinfo  {journal} {J.
  Chem. Phys.}\ }\textbf {\bibinfo {volume} {110}},\ \bibinfo {pages} {8254}
  (\bibinfo {year} {1999})}\BibitemShut {NoStop}%
\bibitem [{\citenamefont {Armstrong}, \citenamefont {Daub},\ and\ \citenamefont
  {Bresme}(2015)}]{Armstrong2015}%
  \BibitemOpen
  \bibfield  {author} {\bibinfo {author} {\bibfnamefont {J.}~\bibnamefont
  {Armstrong}}, \bibinfo {author} {\bibfnamefont {C.~D.}\ \bibnamefont {Daub}},
  \ and\ \bibinfo {author} {\bibfnamefont {F.}~\bibnamefont {Bresme}},\ }\href
  {\doibase 10.1063/1.4927229} {\bibfield  {journal} {\bibinfo  {journal} {J.
  Chem. Phys.}\ }\textbf {\bibinfo {volume} {143}},\ \bibinfo {pages} {036101}
  (\bibinfo {year} {2015})}\BibitemShut {NoStop}%
\bibitem [{\citenamefont {Steinbach}\ and\ \citenamefont
  {Brooks}(1994)}]{Steinbach1994}%
  \BibitemOpen
  \bibfield  {author} {\bibinfo {author} {\bibfnamefont {P.~J.}\ \bibnamefont
  {Steinbach}}\ and\ \bibinfo {author} {\bibfnamefont {B.~R.}\ \bibnamefont
  {Brooks}},\ }\href {\doibase 10.1002/jcc.540150702} {\bibfield  {journal}
  {\bibinfo  {journal} {J. Comput. Chem.}\ }\textbf {\bibinfo {volume} {15}},\
  \bibinfo {pages} {667} (\bibinfo {year} {1994})}\BibitemShut {NoStop}%
\bibitem [{\citenamefont {Zahn}, \citenamefont {Schilling},\ and\ \citenamefont
  {Kast}(2002)}]{Zahn2002}%
  \BibitemOpen
  \bibfield  {author} {\bibinfo {author} {\bibfnamefont {D.}~\bibnamefont
  {Zahn}}, \bibinfo {author} {\bibfnamefont {B.}~\bibnamefont {Schilling}}, \
  and\ \bibinfo {author} {\bibfnamefont {S.~M.}\ \bibnamefont {Kast}},\ }\href
  {\doibase 10.1021/jp025949h} {\bibfield  {journal} {\bibinfo  {journal} {J.
  Chem. Phys. B}\ }\textbf {\bibinfo {volume} {106}},\ \bibinfo {pages} {10725}
  (\bibinfo {year} {2002})}\BibitemShut {NoStop}%
\bibitem [{\citenamefont {Wu}\ and\ \citenamefont {Brooks}(2005)}]{Wu2005}%
  \BibitemOpen
  \bibfield  {author} {\bibinfo {author} {\bibfnamefont {X.}~\bibnamefont
  {Wu}}\ and\ \bibinfo {author} {\bibfnamefont {B.~R.}\ \bibnamefont
  {Brooks}},\ }\href {\doibase 10.1063/1.1836733} {\bibfield  {journal}
  {\bibinfo  {journal} {J. Chem. Phys.}\ }\textbf {\bibinfo {volume} {122}},\
  \bibinfo {pages} {44107} (\bibinfo {year} {2005})}\BibitemShut {NoStop}%
\bibitem [{\citenamefont {Elvira}\ and\ \citenamefont
  {MacDowell}(2014)}]{Elvira2014a}%
  \BibitemOpen
  \bibfield  {author} {\bibinfo {author} {\bibfnamefont {V.~H.}\ \bibnamefont
  {Elvira}}\ and\ \bibinfo {author} {\bibfnamefont {L.~G.}\ \bibnamefont
  {MacDowell}},\ }\href {\doibase 10.1063/1.4898147} {\bibfield  {journal}
  {\bibinfo  {journal} {J. Chem. Phys.}\ }\textbf {\bibinfo {volume} {141}},\
  \bibinfo {pages} {164108} (\bibinfo {year} {2014})}\BibitemShut {NoStop}%
\bibitem [{\citenamefont {Fukuda}, \citenamefont {Yonezawa},\ and\
  \citenamefont {Nakamura}(2011)}]{Fukuda2011}%
  \BibitemOpen
  \bibfield  {author} {\bibinfo {author} {\bibfnamefont {I.}~\bibnamefont
  {Fukuda}}, \bibinfo {author} {\bibfnamefont {Y.}~\bibnamefont {Yonezawa}}, \
  and\ \bibinfo {author} {\bibfnamefont {H.}~\bibnamefont {Nakamura}},\ }\href
  {\doibase 10.1063/1.3582791} {\bibfield  {journal} {\bibinfo  {journal} {J.
  Chem. Phys.}\ }\textbf {\bibinfo {volume} {134}},\ \bibinfo {pages} {164107}
  (\bibinfo {year} {2011})}\BibitemShut {NoStop}%
\bibitem [{\citenamefont {Fukuda}(2013)}]{Fukuda2013}%
  \BibitemOpen
  \bibfield  {author} {\bibinfo {author} {\bibfnamefont {I.}~\bibnamefont
  {Fukuda}},\ }\href {\doibase 10.1063/1.4827055} {\bibfield  {journal}
  {\bibinfo  {journal} {J. Chem. Phys.}\ }\textbf {\bibinfo {volume} {139}},\
  \bibinfo {pages} {174107} (\bibinfo {year} {2013})}\BibitemShut {NoStop}%
\bibitem [{\citenamefont {Lamichhane}, \citenamefont {Gezelter},\ and\
  \citenamefont {Newman}(2014)}]{Lamichhane2014a}%
  \BibitemOpen
  \bibfield  {author} {\bibinfo {author} {\bibfnamefont {M.}~\bibnamefont
  {Lamichhane}}, \bibinfo {author} {\bibfnamefont {J.~D.}\ \bibnamefont
  {Gezelter}}, \ and\ \bibinfo {author} {\bibfnamefont {K.~E.}\ \bibnamefont
  {Newman}},\ }\href {\doibase 10.1063/1.4896627} {\bibfield  {journal}
  {\bibinfo  {journal} {J. Chem. Phys.}\ }\textbf {\bibinfo {volume} {141}},\
  \bibinfo {pages} {134109} (\bibinfo {year} {2014})}\BibitemShut {NoStop}%
\bibitem [{\citenamefont {Chen}\ and\ \citenamefont {Weeks}(2006)}]{Chen2006a}%
  \BibitemOpen
  \bibfield  {author} {\bibinfo {author} {\bibfnamefont {Y.-G.}\ \bibnamefont
  {Chen}}\ and\ \bibinfo {author} {\bibfnamefont {J.~D.}\ \bibnamefont
  {Weeks}},\ }\href {\doibase 10.1073/pnas.0600282103} {\bibfield  {journal}
  {\bibinfo  {journal} {Proc. Natl. Acad. Sci.}\ }\textbf {\bibinfo {volume}
  {103}},\ \bibinfo {pages} {7560} (\bibinfo {year} {2006})}\BibitemShut
  {NoStop}%
\bibitem [{\citenamefont {Fanourgakis}(2015)}]{Fanourgakis2015}%
  \BibitemOpen
  \bibfield  {author} {\bibinfo {author} {\bibfnamefont {G.~S.}\ \bibnamefont
  {Fanourgakis}},\ }\href {\doibase 10.1021/jp510612w} {\bibfield  {journal}
  {\bibinfo  {journal} {J. Phys. Chem. B}\ }\textbf {\bibinfo {volume} {119}},\
  \bibinfo {pages} {1974} (\bibinfo {year} {2015})}\BibitemShut {NoStop}%
\bibitem [{\citenamefont {Darden}, \citenamefont {York},\ and\ \citenamefont
  {Pedersen}(1993)}]{Darden1993}%
  \BibitemOpen
  \bibfield  {author} {\bibinfo {author} {\bibfnamefont {T.}~\bibnamefont
  {Darden}}, \bibinfo {author} {\bibfnamefont {D.}~\bibnamefont {York}}, \ and\
  \bibinfo {author} {\bibfnamefont {L.}~\bibnamefont {Pedersen}},\ }\href
  {\doibase 10.1063/1.464397} {\bibfield  {journal} {\bibinfo  {journal} {J.
  Chem. Phys.}\ }\textbf {\bibinfo {volume} {98}},\ \bibinfo {pages} {10089}
  (\bibinfo {year} {1993})}\BibitemShut {NoStop}%
\bibitem [{\citenamefont {Deserno}\ and\ \citenamefont
  {Holm}(1998)}]{Deserno1998}%
  \BibitemOpen
  \bibfield  {author} {\bibinfo {author} {\bibfnamefont {M.}~\bibnamefont
  {Deserno}}\ and\ \bibinfo {author} {\bibfnamefont {C.}~\bibnamefont {Holm}},\
  }\href {\doibase 10.1063/1.477414} {\bibfield  {journal} {\bibinfo  {journal}
  {J. Chem. Phys.}\ }\textbf {\bibinfo {volume} {109}},\ \bibinfo {pages}
  {7678} (\bibinfo {year} {1998})}\BibitemShut {NoStop}%
\bibitem [{\citenamefont {Neumann}\ and\ \citenamefont
  {Steinhauser}(1980)}]{Neumann1980}%
  \BibitemOpen
  \bibfield  {author} {\bibinfo {author} {\bibfnamefont {M.}~\bibnamefont
  {Neumann}}\ and\ \bibinfo {author} {\bibfnamefont {O.}~\bibnamefont
  {Steinhauser}},\ }\href {\doibase 10.1080/00268978000100361} {\bibfield
  {journal} {\bibinfo  {journal} {Mol. Phys.}\ }\textbf {\bibinfo {volume}
  {39}},\ \bibinfo {pages} {437} (\bibinfo {year} {1980})}\BibitemShut
  {NoStop}%
\bibitem [{\citenamefont {Schreiber}\ and\ \citenamefont
  {Steinhauser}(1992)}]{Schreiber1992}%
  \BibitemOpen
  \bibfield  {author} {\bibinfo {author} {\bibfnamefont {H.}~\bibnamefont
  {Schreiber}}\ and\ \bibinfo {author} {\bibfnamefont {O.}~\bibnamefont
  {Steinhauser}},\ }\href {\doibase 10.1021/bi00140a022} {\bibfield  {journal}
  {\bibinfo  {journal} {Biochemistry}\ }\textbf {\bibinfo {volume} {31}},\
  \bibinfo {pages} {5856} (\bibinfo {year} {1992})}\BibitemShut {NoStop}%
\bibitem [{\citenamefont {Feller}\ \emph {et~al.}(1996)\citenamefont {Feller},
  \citenamefont {Pastor}, \citenamefont {Rojnuckarin}, \citenamefont {Bogusz},\
  and\ \citenamefont {Brooks}}]{Feller1996}%
  \BibitemOpen
  \bibfield  {author} {\bibinfo {author} {\bibfnamefont {S.~E.}\ \bibnamefont
  {Feller}}, \bibinfo {author} {\bibfnamefont {R.~W.}\ \bibnamefont {Pastor}},
  \bibinfo {author} {\bibfnamefont {A.}~\bibnamefont {Rojnuckarin}}, \bibinfo
  {author} {\bibfnamefont {S.}~\bibnamefont {Bogusz}}, \ and\ \bibinfo {author}
  {\bibfnamefont {B.~R.}\ \bibnamefont {Brooks}},\ }\href {\doibase
  10.1021/jp9614658} {\bibfield  {journal} {\bibinfo  {journal} {J. Phys.
  Chem.}\ }\textbf {\bibinfo {volume} {100}},\ \bibinfo {pages} {17011}
  (\bibinfo {year} {1996})}\BibitemShut {NoStop}%
\bibitem [{\citenamefont {Rodgers}\ and\ \citenamefont
  {Weeks}(2008{\natexlab{a}})}]{Rodgers2008a}%
  \BibitemOpen
  \bibfield  {author} {\bibinfo {author} {\bibfnamefont {J.~M.}\ \bibnamefont
  {Rodgers}}\ and\ \bibinfo {author} {\bibfnamefont {J.~D.}\ \bibnamefont
  {Weeks}},\ }\href {\doibase 10.1073/pnas.0807623105} {\bibfield  {journal}
  {\bibinfo  {journal} {Proc. Natl. Acad. Sci.}\ }\textbf {\bibinfo {volume}
  {105}},\ \bibinfo {pages} {19136} (\bibinfo {year}
  {2008}{\natexlab{a}})}\BibitemShut {NoStop}%
\bibitem [{\citenamefont {Spohr}(1997)}]{Spohr1997}%
  \BibitemOpen
  \bibfield  {author} {\bibinfo {author} {\bibfnamefont {E.}~\bibnamefont
  {Spohr}},\ }\href {\doibase 10.1063/1.474295} {\bibfield  {journal} {\bibinfo
   {journal} {J. Chem. Phys.}\ }\textbf {\bibinfo {volume} {107}},\ \bibinfo
  {pages} {6342} (\bibinfo {year} {1997})}\BibitemShut {NoStop}%
\bibitem [{\citenamefont {{Van der Spoel}}\ and\ \citenamefont {{Van
  Maaren}}(2006)}]{VanDerSpoel2006}%
  \BibitemOpen
  \bibfield  {author} {\bibinfo {author} {\bibfnamefont {D.}~\bibnamefont {{Van
  der Spoel}}}\ and\ \bibinfo {author} {\bibfnamefont {P.~J.}\ \bibnamefont
  {{Van Maaren}}},\ }\href {\doibase 10.1021/ct0502256} {\bibfield  {journal}
  {\bibinfo  {journal} {J. Chem. Theory Comput.}\ }\textbf {\bibinfo {volume}
  {2}},\ \bibinfo {pages} {1} (\bibinfo {year} {2006})}\BibitemShut {NoStop}%
\bibitem [{\citenamefont {Cisneros}\ \emph {et~al.}(2014)\citenamefont
  {Cisneros}, \citenamefont {Karttunen}, \citenamefont {Ren},\ and\
  \citenamefont {Sagui}}]{Cisneros2014}%
  \BibitemOpen
  \bibfield  {author} {\bibinfo {author} {\bibfnamefont {G.~A.}\ \bibnamefont
  {Cisneros}}, \bibinfo {author} {\bibfnamefont {M.}~\bibnamefont {Karttunen}},
  \bibinfo {author} {\bibfnamefont {P.}~\bibnamefont {Ren}}, \ and\ \bibinfo
  {author} {\bibfnamefont {C.}~\bibnamefont {Sagui}},\ }\href {\doibase
  10.1021/cr300461d} {\bibfield  {journal} {\bibinfo  {journal} {Chem. Rev.}\
  }\textbf {\bibinfo {volume} {114}},\ \bibinfo {pages} {779} (\bibinfo {year}
  {2014})}\BibitemShut {NoStop}%
\bibitem [{\citenamefont {Muscatello}\ and\ \citenamefont
  {Bresme}(2011)}]{Muscatello2011}%
  \BibitemOpen
  \bibfield  {author} {\bibinfo {author} {\bibfnamefont {J.}~\bibnamefont
  {Muscatello}}\ and\ \bibinfo {author} {\bibfnamefont {F.}~\bibnamefont
  {Bresme}},\ }\href {\doibase 10.1063/1.3670965} {\bibfield  {journal}
  {\bibinfo  {journal} {J. Chem. Phys.}\ }\textbf {\bibinfo {volume} {135}},\
  \bibinfo {pages} {234111} (\bibinfo {year} {2011})}\BibitemShut {NoStop}%
\bibitem [{\citenamefont {{No{\'{e}} Mendoza}}\ \emph
  {et~al.}(2008)\citenamefont {{No{\'{e}} Mendoza}}, \citenamefont
  {L{\'{o}}pez-Lemus}, \citenamefont {Chapela},\ and\ \citenamefont
  {Alejandre}}]{Mendoza2008}%
  \BibitemOpen
  \bibfield  {author} {\bibinfo {author} {\bibfnamefont {F.}~\bibnamefont
  {{No{\'{e}} Mendoza}}}, \bibinfo {author} {\bibfnamefont {J.}~\bibnamefont
  {L{\'{o}}pez-Lemus}}, \bibinfo {author} {\bibfnamefont {G.~A.}\ \bibnamefont
  {Chapela}}, \ and\ \bibinfo {author} {\bibfnamefont {J.}~\bibnamefont
  {Alejandre}},\ }\href {\doibase 10.1063/1.2948951} {\bibfield  {journal}
  {\bibinfo  {journal} {J. Chem. Phys.}\ }\textbf {\bibinfo {volume} {129}},\
  \bibinfo {pages} {024706} (\bibinfo {year} {2008})}\BibitemShut {NoStop}%
\bibitem [{\citenamefont {Takahashi}, \citenamefont {Narumi},\ and\
  \citenamefont {Yasuoka}(2011)}]{Takahashi2011}%
  \BibitemOpen
  \bibfield  {author} {\bibinfo {author} {\bibfnamefont {K.~Z.}\ \bibnamefont
  {Takahashi}}, \bibinfo {author} {\bibfnamefont {T.}~\bibnamefont {Narumi}}, \
  and\ \bibinfo {author} {\bibfnamefont {K.}~\bibnamefont {Yasuoka}},\ }\href
  {\doibase 10.1063/1.3578473} {\bibfield  {journal} {\bibinfo  {journal} {J.
  Chem. Phys.}\ }\textbf {\bibinfo {volume} {134}},\ \bibinfo {pages} {174112}
  (\bibinfo {year} {2011})}\BibitemShut {NoStop}%
\bibitem [{\citenamefont {Rodgers}\ and\ \citenamefont
  {Weeks}(2008{\natexlab{b}})}]{Rodgers2008}%
  \BibitemOpen
  \bibfield  {author} {\bibinfo {author} {\bibfnamefont {J.~M.}\ \bibnamefont
  {Rodgers}}\ and\ \bibinfo {author} {\bibfnamefont {J.~D.}\ \bibnamefont
  {Weeks}},\ }\href {\doibase 10.1088/0953-8984/20/49/494206} {\bibfield
  {journal} {\bibinfo  {journal} {J. Phys.: Condens. Matter}\ }\textbf
  {\bibinfo {volume} {20}},\ \bibinfo {pages} {494206} (\bibinfo {year}
  {2008}{\natexlab{b}})}\BibitemShut {NoStop}%
\bibitem [{\citenamefont {De~Leeuw}, \citenamefont {Perram},\ and\
  \citenamefont {Smith}(1980)}]{DeLeeuw1980}%
  \BibitemOpen
  \bibfield  {author} {\bibinfo {author} {\bibfnamefont {S.~W.}\ \bibnamefont
  {De~Leeuw}}, \bibinfo {author} {\bibfnamefont {J.~W.}\ \bibnamefont
  {Perram}}, \ and\ \bibinfo {author} {\bibfnamefont {E.~R.}\ \bibnamefont
  {Smith}},\ }\href {\doibase 10.1098/rspa.1980.0135} {\bibfield  {journal}
  {\bibinfo  {journal} {Proc. R. Soc. London, Ser. A}\ }\textbf {\bibinfo
  {volume} {373}},\ \bibinfo {pages} {27} (\bibinfo {year} {1980})}\BibitemShut
  {NoStop}%
\bibitem [{\citenamefont {Neumann}, \citenamefont {Steinhauser},\ and\
  \citenamefont {Pawley}(1984)}]{Neumann1984}%
  \BibitemOpen
  \bibfield  {author} {\bibinfo {author} {\bibfnamefont {M.}~\bibnamefont
  {Neumann}}, \bibinfo {author} {\bibfnamefont {O.}~\bibnamefont
  {Steinhauser}}, \ and\ \bibinfo {author} {\bibfnamefont {G.~S.}\ \bibnamefont
  {Pawley}},\ }\href {\doibase 10.1080/00268978400101081} {\bibfield  {journal}
  {\bibinfo  {journal} {Mol. Phys.}\ }\textbf {\bibinfo {volume} {52}},\
  \bibinfo {pages} {97} (\bibinfo {year} {1984})}\BibitemShut {NoStop}%
\bibitem [{\citenamefont {Wick}, \citenamefont {Dang},\ and\ \citenamefont
  {Jungwirth}(2006)}]{Wick2006}%
  \BibitemOpen
  \bibfield  {author} {\bibinfo {author} {\bibfnamefont {C.~D.}\ \bibnamefont
  {Wick}}, \bibinfo {author} {\bibfnamefont {L.~X.}\ \bibnamefont {Dang}}, \
  and\ \bibinfo {author} {\bibfnamefont {P.}~\bibnamefont {Jungwirth}},\ }\href
  {\doibase 10.1063/1.2218840} {\bibfield  {journal} {\bibinfo  {journal} {J.
  Chem. Phys.}\ }\textbf {\bibinfo {volume} {125}},\ \bibinfo {pages} {024706}
  (\bibinfo {year} {2006})}\BibitemShut {NoStop}%
\bibitem [{\citenamefont {Wilson}, \citenamefont {Pohorille},\ and\
  \citenamefont {Pratt}(1988)}]{Wilson1988}%
  \BibitemOpen
  \bibfield  {author} {\bibinfo {author} {\bibfnamefont {M.~A.}\ \bibnamefont
  {Wilson}}, \bibinfo {author} {\bibfnamefont {A.}~\bibnamefont {Pohorille}}, \
  and\ \bibinfo {author} {\bibfnamefont {L.~R.}\ \bibnamefont {Pratt}},\ }\href
  {\doibase 10.1063/1.453923} {\bibfield  {journal} {\bibinfo  {journal} {J.
  Chem. Phys.}\ }\textbf {\bibinfo {volume} {88}},\ \bibinfo {pages} {3281}
  (\bibinfo {year} {1988})}\BibitemShut {NoStop}%
\bibitem [{\citenamefont {Yeh}\ and\ \citenamefont
  {Berkowitz}(1999)}]{Yeh1999}%
  \BibitemOpen
  \bibfield  {author} {\bibinfo {author} {\bibfnamefont {I.-C.}\ \bibnamefont
  {Yeh}}\ and\ \bibinfo {author} {\bibfnamefont {M.~L.}\ \bibnamefont
  {Berkowitz}},\ }\href {\doibase 10.1063/1.479595} {\bibfield  {journal} {\bibinfo  {journal} {J.
  Chem. Phys.}\ }\textbf {\bibinfo {volume} {111}},\ \bibinfo {pages} {3155} (\bibinfo {year}
  {1999})}\BibitemShut {NoStop}%
\bibitem [{\citenamefont {Glosli}\ and\ \citenamefont
  {Philpott}(1996)}]{Glosli1996}%
  \BibitemOpen
  \bibfield  {author} {\bibinfo {author} {\bibfnamefont {J.~N.}\ \bibnamefont
  {Glosli}}\ and\ \bibinfo {author} {\bibfnamefont {M.~R.}\ \bibnamefont
  {Philpott}},\ }\href {\doibase 10.1016/0013-4686(96)00046-1} {\bibfield
  {journal} {\bibinfo  {journal} {Electrochim. Acta}\ }\textbf {\bibinfo
  {volume} {41}},\ \bibinfo {pages} {2145} (\bibinfo {year}
  {1996})}\BibitemShut {NoStop}%
\bibitem [{\citenamefont {Frenkel}\ and\ \citenamefont
  {Smit}(2002)}]{Frenkel2002}%
  \BibitemOpen
  \bibfield  {author} {\bibinfo {author} {\bibfnamefont {D.}~\bibnamefont
  {Frenkel}}\ and\ \bibinfo {author} {\bibfnamefont {B.}~\bibnamefont {Smit}},\
  }\href@noop {} {\emph {\bibinfo {title} {{Understanding Molecular
  Simulation}}}},\ \bibinfo {edition} {2nd}\ ed.\ (\bibinfo  {publisher}
  {Academic Press},\ \bibinfo {address} {San Diego},\ \bibinfo {year}
  {2002})\BibitemShut {NoStop}%
\bibitem [{\citenamefont {Hummer}\ \emph {et~al.}(1999)\citenamefont {Hummer},
  \citenamefont {Pratt}, \citenamefont {Garc\'{\i}a},\ and\ \citenamefont
  {Neumann}}]{Hummer1999b}%
  \BibitemOpen
  \bibfield  {author} {\bibinfo {author} {\bibfnamefont {G.}~\bibnamefont
  {Hummer}}, \bibinfo {author} {\bibfnamefont {L.~R.}\ \bibnamefont {Pratt}},
  \bibinfo {author} {\bibfnamefont {A.~E.}\ \bibnamefont {Garc\'{\i}a}}, \ and\
  \bibinfo {author} {\bibfnamefont {M.}~\bibnamefont {Neumann}},\ }\href
  {\doibase 10.1063/1.1301522} {\bibfield  {journal} {\bibinfo  {journal} {AIP
  Conf. Proc.}\ }\textbf {\bibinfo {volume} {492}},\ \bibinfo {pages} {84}
  (\bibinfo {year} {1999})}\BibitemShut {NoStop}%
\bibitem [{\citenamefont {Hummer}, \citenamefont {Gr{\o}nbech-Jensen},\ and\
  \citenamefont {Neumann}(1998)}]{Hummer1998}%
  \BibitemOpen
  \bibfield  {author} {\bibinfo {author} {\bibfnamefont {G.}~\bibnamefont
  {Hummer}}, \bibinfo {author} {\bibfnamefont {N.}~\bibnamefont
  {Gr{\o}nbech-Jensen}}, \ and\ \bibinfo {author} {\bibfnamefont
  {M.}~\bibnamefont {Neumann}},\ }\href {\doibase 10.1063/1.476834} {\bibfield
  {journal} {\bibinfo  {journal} {J. Chem. Phys.}\ }\textbf {\bibinfo {volume}
  {109}},\ \bibinfo {pages} {2791} (\bibinfo {year} {1998})}\BibitemShut
  {NoStop}%
\bibitem [{\citenamefont {Berendsen}, \citenamefont {Grigera},\ and\
  \citenamefont {Straatsma}(1987)}]{Berendsen1987}%
  \BibitemOpen
  \bibfield  {author} {\bibinfo {author} {\bibfnamefont {H.~J.~C.}\
  \bibnamefont {Berendsen}}, \bibinfo {author} {\bibfnamefont {J.~R.}\
  \bibnamefont {Grigera}}, \ and\ \bibinfo {author} {\bibfnamefont {T.~P.}\
  \bibnamefont {Straatsma}},\ }\href {\doibase 10.1021/j100308a038} {\bibfield
  {journal} {\bibinfo  {journal} {J. Phys. Chem.}\ }\textbf {\bibinfo {volume}
  {91}},\ \bibinfo {pages} {6269} (\bibinfo {year} {1987})}\BibitemShut
  {NoStop}%
\bibitem [{\citenamefont {Wilson}, \citenamefont {Pohorille},\ and\
  \citenamefont {Pratt}(1989)}]{Wilson1989}%
  \BibitemOpen
  \bibfield  {author} {\bibinfo {author} {\bibfnamefont {M.~A.}\ \bibnamefont
  {Wilson}}, \bibinfo {author} {\bibfnamefont {A.}~\bibnamefont {Pohorille}}, \
  and\ \bibinfo {author} {\bibfnamefont {L.~R.}\ \bibnamefont {Pratt}},\ }\href
  {\doibase 10.1063/1.453919} {\bibfield  {journal} {\bibinfo  {journal} {J.
  Chem. Phys.}\ }\textbf {\bibinfo {volume} {90}},\ \bibinfo {pages} {5211}
  (\bibinfo {year} {1989})}\BibitemShut {NoStop}%
\bibitem [{\citenamefont {Smith}(1998)}]{Smith1998}%
  \BibitemOpen
  \bibfield  {author} {\bibinfo {author} {\bibfnamefont {W.}~\bibnamefont
  {Smith}},\ }\href@noop {} {\bibfield  {journal} {\bibinfo  {journal} {CCP5
  Newsletter}\ }\textbf {\bibinfo {volume} {46}},\ \bibinfo {pages} {18}
  (\bibinfo {year} {1998})}\BibitemShut {NoStop}%
\bibitem [{\citenamefont {Jackson}(1998)}]{Jackson1998}%
  \BibitemOpen
  \bibfield  {author} {\bibinfo {author} {\bibfnamefont {J.~D.}\ \bibnamefont
  {Jackson}},\ }\href@noop {} {\emph {\bibinfo {title} {{Classical
  electrodynamics}}}},\ \bibinfo {edition} {3rd}\ ed.\ (\bibinfo  {publisher}
  {Wiley},\ \bibinfo {year} {1998})\BibitemShut
  {NoStop}%
\bibitem [{\citenamefont {Spaldin}(2012)}]{Spaldin2012}%
  \BibitemOpen
  \bibfield  {author} {\bibinfo {author} {\bibfnamefont {N.~A.}\ \bibnamefont
  {Spaldin}},\ }\href {\doibase 10.1016/j.jssc.2012.05.010} {\bibfield
  {journal} {\bibinfo  {journal} {J. Solid State Chem.}\ }\textbf {\bibinfo
  {volume} {195}},\ \bibinfo {pages} {2} (\bibinfo {year} {2012})}\BibitemShut
  {NoStop}%
\bibitem [{\citenamefont {Plimpton}(1995)}]{Plimpton1995}%
  \BibitemOpen
  \bibfield  {author} {\bibinfo {author} {\bibfnamefont {S.}~\bibnamefont
  {Plimpton}},\ }\href
  {http://www.sciencedirect.com/science/article/pii/S002199918571039X}
  {\bibfield  {journal} {\bibinfo  {journal} {J. Comput. Phys.}\ }\textbf
  {\bibinfo {volume} {117}},\ \bibinfo {pages} {1} (\bibinfo {year}
  {1995})}\BibitemShut {NoStop}%
\bibitem [{\citenamefont {Wirnsberger}, \citenamefont {Frenkel},\ and\
  \citenamefont {Dellago}(2015)}]{Wirnsberger2015}%
  \BibitemOpen
  \bibfield  {author} {\bibinfo {author} {\bibfnamefont {P.}~\bibnamefont
  {Wirnsberger}}, \bibinfo {author} {\bibfnamefont {D.}~\bibnamefont
  {Frenkel}}, \ and\ \bibinfo {author} {\bibfnamefont {C.}~\bibnamefont
  {Dellago}},\ }\href {\doibase 10.1063/1.4931597} {\bibfield  {journal}
  {\bibinfo  {journal} {J. Chem. Phys.}\ }\textbf {\bibinfo {volume} {143}},\
  \bibinfo {pages} {124104} (\bibinfo {year} {2015})}\BibitemShut {NoStop}%
\bibitem [{\citenamefont {Swope}\ \emph {et~al.}(1982)\citenamefont {Swope},
  \citenamefont {Andersen}, \citenamefont {Berens},\ and\ \citenamefont
  {Wilson}}]{Swope1982a}%
  \BibitemOpen
  \bibfield  {author} {\bibinfo {author} {\bibfnamefont {W.~C.}\ \bibnamefont
  {Swope}}, \bibinfo {author} {\bibfnamefont {H.~C.}\ \bibnamefont {Andersen}},
  \bibinfo {author} {\bibfnamefont {P.~H.}\ \bibnamefont {Berens}}, \ and\
  \bibinfo {author} {\bibfnamefont {K.~R.}\ \bibnamefont {Wilson}},\ }\href
  {\doibase 10.1063/1.442716} {\bibfield  {journal} {\bibinfo  {journal} {J.
  Chem. Phys.}\ }\textbf {\bibinfo {volume} {76}},\ \bibinfo {pages} {637}
  (\bibinfo {year} {1982})}\BibitemShut {NoStop}%
\bibitem [{\citenamefont {Nos\'{e}}(1984)}]{Nose1984}%
  \BibitemOpen
  \bibfield  {author} {\bibinfo {author} {\bibfnamefont {S.}~\bibnamefont
  {Nos\'{e}}},\ }\href {\doibase 10.1063/1.447334} {\bibfield  {journal}
  {\bibinfo  {journal} {J. Chem. Phys.}\ }\textbf {\bibinfo {volume} {81}},\
  \bibinfo {pages} {511} (\bibinfo {year} {1984})}\BibitemShut {NoStop}%
\bibitem [{\citenamefont {Hoover}(1985)}]{Hoover1985}%
  \BibitemOpen
  \bibfield  {author} {\bibinfo {author} {\bibfnamefont {W.~G.}\ \bibnamefont
  {Hoover}},\ }\href {\doibase 10.1103/PhysRevA.31.1695} {\bibfield  {journal}
  {\bibinfo  {journal} {Phys. Rev. A}\ }\textbf {\bibinfo {volume} {31}},\
  \bibinfo {pages} {1695} (\bibinfo {year} {1985})}\BibitemShut {NoStop}%
\bibitem [{\citenamefont {Yeh}\ and\ \citenamefont
  {Wallqvist}(2011)}]{Yeh2011}%
  \BibitemOpen
  \bibfield  {author} {\bibinfo {author} {\bibfnamefont {I.-C.}\ \bibnamefont
  {Yeh}}\ and\ \bibinfo {author} {\bibfnamefont {A.}~\bibnamefont
  {Wallqvist}},\ }\href {\doibase 10.1063/1.3548836} {\bibfield  {journal}
  {\bibinfo  {journal} {J. Chem. Phys.}\ }\textbf {\bibinfo {volume} {134}},\
  \bibinfo {pages} {055109} (\bibinfo {year} {2011})}\BibitemShut {NoStop}%
\bibitem [{\citenamefont {Neumann}(1986{\natexlab{a}})}]{Neumann1986a}%
  \BibitemOpen
  \bibfield  {author} {\bibinfo {author} {\bibfnamefont {M.}~\bibnamefont
  {Neumann}},\ }\href {\doibase 10.1080/00268978600100081} {\bibfield
  {journal} {\bibinfo  {journal} {Mol. Phys.}\ }\textbf {\bibinfo {volume}
  {57}},\ \bibinfo {pages} {97} (\bibinfo {year}
  {1986}{\natexlab{a}})}\BibitemShut {NoStop}%
\bibitem [{\citenamefont {B\"ottcher}(1973)}]{Boettcher1973}%
  \BibitemOpen
  \bibfield  {author} {\bibinfo {author} {\bibfnamefont {C.~J.~F.}\
  \bibnamefont {B\"ottcher}},\ }\href@noop {} {\emph {\bibinfo {title}
  {{Dielectrics in static fields}}}},\ \bibinfo {edition} {2nd}\ ed.\ (\bibinfo
   {publisher} {Elsevier Science},\ \bibinfo {year} {1973})
  \BibitemShut {NoStop}%
\bibitem [{\citenamefont {Sokhan}\ and\ \citenamefont
  {Tildesley}(1997)}]{Sokhan1997}%
  \BibitemOpen
  \bibfield  {author} {\bibinfo {author} {\bibfnamefont {V.~P.}\ \bibnamefont
  {Sokhan}}\ and\ \bibinfo {author} {\bibfnamefont {D.~J.}\ \bibnamefont
  {Tildesley}},\ }\href {\doibase 10.1080/002689797169916} {\bibfield
  {journal} {\bibinfo  {journal} {Mol. Phys.}\ }\textbf {\bibinfo {volume}
  {92}},\ \bibinfo {pages} {625} (\bibinfo {year} {1997})}\BibitemShut
  {NoStop}%
\bibitem [{\citenamefont {Stengel}, \citenamefont {Spaldin},\ and\
  \citenamefont {Vanderbilt}(2008)}]{Stengel2008}%
  \BibitemOpen
  \bibfield  {author} {\bibinfo {author} {\bibfnamefont {M.}~\bibnamefont
  {Stengel}}, \bibinfo {author} {\bibfnamefont {N.~A.}\ \bibnamefont
  {Spaldin}}, \ and\ \bibinfo {author} {\bibfnamefont {D.}~\bibnamefont
  {Vanderbilt}},\ }\href {\doibase 10.1038/nphys1185} {\bibfield  {journal}
  {\bibinfo  {journal} {Nat. Phys.}\ }\textbf {\bibinfo {volume} {5}},\
  \bibinfo {pages} {304} (\bibinfo {year} {2009})}\BibitemShut {NoStop}%
\bibitem [{\citenamefont {Zhang}\ and\ \citenamefont
  {Sprik}(2016)}]{Zhang2016}%
  \BibitemOpen
  \bibfield  {author} {\bibinfo {author} {\bibfnamefont {C.}~\bibnamefont
  {Zhang}}\ and\ \bibinfo {author} {\bibfnamefont {M.}~\bibnamefont {Sprik}},\
  }\href {\doibase 10.1103/PhysRevB.93.144201} {\bibfield  {journal} {\bibinfo
  {journal} {Phys. Rev. B}\ }\textbf {\bibinfo {volume} {93}},\ \bibinfo
  {pages} {144201} (\bibinfo {year} {2016})}\BibitemShut {NoStop}%
\bibitem [{\citenamefont {Neumann}(1986{\natexlab{b}})}]{Neumann1986}%
  \BibitemOpen
  \bibfield  {author} {\bibinfo {author} {\bibfnamefont {M.}~\bibnamefont
  {Neumann}},\ }\href {\doibase 10.1063/1.451198} {\bibfield  {journal}
  {\bibinfo  {journal} {J. Chem. Phys.}\ }\textbf {\bibinfo {volume} {85}},\
  \bibinfo {pages} {1567} (\bibinfo {year} {1986}{\natexlab{b}})}\BibitemShut
  {NoStop}%
\end{thebibliography}
\end{document}